\documentclass[12pt,a4paper,final]{article}
\usepackage{times}
\usepackage{a4wide}  
\usepackage{amsfonts}
\usepackage{amssymb}
\usepackage{amsmath}
\usepackage{bbm}
\usepackage{booktabs} 
\usepackage[notref,notcite]{showkeys}
\usepackage{ifpdf}
\ifpdf
\usepackage[pdftex,unicode,implicit]{hyperref}
\hypersetup{%
  pdftitle    = {Black holes and equivariant charge vectors in N=2,d=4 supergravity},
  pdfkeywords = {black hole, equivariant vector, supergravity},
  pdfauthor   = {Pablo Bueno, Pietro Galli, Patrick Meessen and Tomas Ortin},
  plainpages  = true,
  colorlinks  = true,
  citecolor   = blue,
  urlcolor    = red,
  linkcolor   = black
}
\newcommand{\hepth}[1]{{\tt
\href{http://www.arXiv.org/abs/hep-th/#1}{hep-th/#1}}}

\newcommand{\arxiv}[1]{{\tt
\href{http://www.arXiv.org/abs/#1}{#1}}}
\else
  \usepackage[dvips]{graphicx}
  \usepackage[unicode,implicit]{hyperref}
  \newcommand{\hepth}[1]{{\tt hep-th/#1}}

  \newcommand{\arxiv}[1]{{\tt arXiv:#1}}
\fi
\usepackage{tikz}\newcommand{\FPAUO}[2]{
\tikz[scale=.13,
         Uniovi/.style={color=green!51!blue, fill=green!51!blue}
 ] {
 \fill[Uniovi] (0,0) circle (10);
 \fill[white] (0,7) circle (1.5);
 \draw[Uniovi] (-2,7.5) rectangle (2,5.5);
 \fill[white] (-0.3,6.6) rectangle (0.3,0);   
 \fill[white] ( -0.9,6.2) rectangle (.9 ,5.6);
 \fill[white] (-1.4, 5.2) rectangle (1.4, 4.6);
 \fill[white] (0,0) ellipse (3.5 and 4);
 \fill[Uniovi] (-2.5,0.3) rectangle (2.5,-0.3);
 \fill[Uniovi] (-2,2.3) rectangle (2,1.7);
 \fill[Uniovi] (-2,-2.3) rectangle (2,-1.7);
 \fill[white] (-4.5,5.5) rectangle (-2.7,4.9);
 \fill[white] (-3.9,6.1) rectangle (-3.3,4.3);
 \fill[white] (4.5,5.5) rectangle (2.7,4.9);
 \fill[white] (3.9,6.1) rectangle (3.3,4.3);
 \foreach \x in { 0,..., 3 }
   \foreach \y in { 0,...,\x}
    {
     \fill[white] (-6-\x*0.7+\y*1.4,3.5-\x *1.97) -- (-5.6-\x*0.7+\y*1.4,2.4-\x *1.97) -- (-6.4-\x*0.7+\y*1.4,2.4-\x *1.97) -- cycle;
     \fill[white] (6-\x*0.7+\y*1.4,3.5-\x *1.97) -- (5.6-\x*0.7+\y*1.4,2.4-\x *1.97) -- (6.4-\x*0.7+\y*1.4,2.4-\x *1.97) -- cycle;
   };
 \draw (0,-6) node[
                               text centered, 
                               color=white, 
                               font={\fontsize{8}{4}\sffamily\selectfont}
                             ] {FPAUO-#1/#2};
}} 
\makeatletter
\@addtoreset{equation}{section}
\makeatother

\begin{document}

\begin{flushright}
\small
\FPAUO{13}{04}\\
IFIC/13-18\\
IFT-UAM/CSIC-13-031\\
May 23\textsuperscript{rd}, 2013\\
\normalsize
\end{flushright}

\vspace{.5cm}

\begin{center}

{\Large {\bf Black holes and equivariant charge vectors}}\\[.5cm] 
{\Large {\bf  in $\mathcal{N}=2,d=4$ supergravity}}

\vspace{1cm}

\renewcommand{\thefootnote}{\alph{footnote}}
{\sl\large  Pablo Bueno$^{\heartsuit}$}
\footnote{E-mail: {\tt pab.bueno [at] estudiante.uam.es}},
{\sl\large Pietro Galli$^{\clubsuit}$}
\footnote{E-mail: {\tt Pietro.Galli [at] ific.uv.es}},
{\sl\large Patrick Meessen$^{\spadesuit}$}
\footnote{E-mail: {\tt meesenpatrick [at] uniovi.es}}
{\sl\large and Tom\'{a}s Ort\'{\i}n$^{\heartsuit}$}
\footnote{E-mail: {\tt Tomas.Ortin [at] csic.es}}.

\setcounter{footnote}{0}
\renewcommand{\thefootnote}{\arabic{footnote}}

\vspace{1cm}

${}^{\clubsuit}${\it Departament de F\'{\i}sica Te\`orica and IFIC (CSIC-UVEG),
Universitat de Val\`encia,\\
C/ Dr.~Moliner, 50, 46100 Burjassot (Val\`encia), Spain}\\

\vspace{.2cm}

${}^{\spadesuit}${\it HEP Theory Group, Departamento de F\'{\i}sica, Universidad de Oviedo\\
  Avda.~Calvo Sotelo s/n, 33007 Oviedo, Spain}\\

\vspace{.2cm}

${}^{\heartsuit}${\it Instituto de F\'{\i}sica Te\'orica UAM/CSIC\\
C/ Nicol\'as Cabrera, 13--15,  C.U.~Cantoblanco, 28049 Madrid, Spain}\\

\vspace{.5cm}


{\bf Abstract}

\end{center}

\begin{quotation}\small
  We extend previous investigations on the construction of extremal
  supersymmetric and non-supersymmetric solutions in the H-FGK formalism to
  \textit{unconventional} solutions with anharmonic terms. We show how the use
  of \textit{fake} charge vectors equivariant under duality transformations
  simplifies and clarifies the task of identification of the attractors
  of the theory.
\end{quotation}

\newpage
\pagestyle{plain}

\tableofcontents

\section*{Introduction}

The intensive search for black-hole solutions of supergravity theories
over the last 25 years has been a very rewarding one in respect to the
supersymmetric (also known as BPS in the literature, even if this
concept is not equivalent, but wider) ones. Even though the existence
of extremal non-supersymmetric black holes was discovered long time
ago \cite{Khuri:1995xq,Ortin:1997yn} and we know that they are subject
to the same attractor mechanism as the supersymmetric ones
\cite{Ferrara:1997tw}, only a few general families of solutions have
been constructed for some classes of theories \cite{Bossard:2012xs}
and we are still far from having a complete understanding of their
structure and general properties. The situation w.r.t.~non-extremal
solutions, which some of us studied recently in
\cite{Galli:2011fq,Meessen:2011bd,Meessen:2012su,Bueno:2012jc,Galli:2012pt}
is even worse: even if all extremal black-hole solutions may be
deformed ({\em i.e.\/} heated up) to a non-extremal one, then we do
not know the non-extremal deformations of many of them; in general we
don't know whether there are obstructions to such a deformation and
what they are.  We also don't know whether, in each theory, there is
only one family of non-extremal black-hole solutions from which all
the extremal ones can be obtained by taken the appropriate limits,
such as it happens in the few models studied so far
\cite{Kallosh:1992ii,LozanoTellechea:1999my,Galli:2011fq,Meessen:2011bd,Meessen:2012su}.
The (stringy) non-extremal black hole landscape is a largely uncharted
territory.

It is clear that to answer these questions new tools are needed since the
first-order equations associated to unbroken supersymmetry are of no help here
and the second-order equations of motion of the FGK effective action
\cite{Ferrara:1997tw} are still very hard to solve. Several approaches have
been proposed to this end. For instance, it has been shown that in general one can
construct first-order flow equations for extremal non-supersymmetric and
non-extremal black holes
Refs.~\cite{Miller:2006ay,Janssen:2007rc,Ceresole:2007wx,Andrianopoli:2007gt,Perz:2008kh,Andrianopoli:2009je,Galli:2009bj,Galli:2010mg,Chemissany:2010zp}
and many such equations have been constructed. From them one can extract
interesting information about the near-horizon and spacelike infinity limits
(whence about the entropy and mass of the solutions), but in
practice these equations are obtained when the solutions are already known,
which somewhat diminishes their usefulness.

The most common approach to the search of stationary black-hole solutions,
pioneered in Ref.~\cite{Breitenlohner:1987dg}, consists in the dimensional
reduction over the time direction. For 4-dimensional theories, this results
in a 3-dimensional theory consisting of a non-linear $\sigma$-model coupled
to gravity (in 3 dimensions the vector fields can be dualized into scalars).\footnote{
Further assumptions (staticity plus an ansatz for the
3-dimensional metric) lead to the FGK effective action with its characteristic
effective \textit{black-hole potential} \cite{Ferrara:1997tw}.} When the $\sigma$-model
corresponds to a homogeneous space one can show that the system is integrable
and use the standard techniques to classify and obtain explicit black-hole
solutions, see {\em e.g.\/} \cite{Bergshoeff:2008be}. This approach has been quite a
successful one, but for the moment it has not provided complete answers to
the above questions.

More recently, a new approach for the 4- and 5-dimensional $\mathcal{N}=2,d=4$
supergravity theories coupled to $n_{V}$ vector supermultiplets has been
introduced in Ref.~\cite{Meessen:2011aa}\footnote{A closely-related approach
  has been proposed in
  Ref.~\cite{Mohaupt:2009iq,Mohaupt:2010fk,Mohaupt:2011aa}.}. This approach,
dubbed the \textit{H-FGK formalism}, consists in using a convenient set of
variables in the FGK effective action.  These variables arise naturally in the
supersymmetric cases \cite{Gauntlett:2004qy,Meessen:2006tu}, but it has been
shown that they can be used in more general (but always stationary) cases.
The main virtue of the new variables, when compared to the scalar fields
present in the FGK effective action, is that they transform linearly under the
duality group (embedded in $\mathrm{Sp}(2n_{V}+2;\mathbb{R})$ in the $d=4$ case and in
$\mathrm{SO}(n_{V}+1)$ in $d=5$ case). 

In previous works
\cite{Meessen:2012su,Galli:2012pt,Bueno:2012jc,Bueno:2013psa} we have
investigated the description of the simplest families of solutions
(that we will call \textit{conventional} in
Section~\ref{sec-conventional}) for which the $H$-variables are
harmonic functions (in the extremal case) or linear combinations of
hyperbolic sines and cosines (in the non-extremal case).  We have also
studied some general features of the formalism, like the invariance of
the effective action under local Freudenthal duality rotations
\cite{Galli:2012ji}, but thus far we have not exploited the main
feature of the $H$-variables, namely the linear equivariance under
duality transformations of the charges and moduli that characterize a
given solution.

Our main goal in this paper is to study this aspect of the formalism and show
how to exploit the requirement of linear equivariance in order to find attractors and
construct explicit extremal solutions in some already well-studied models: the
axidilaton and the $\overline{\mathbb{CP}}^{n}$ models. We also want to make
progress towards answering the questions posed at the beginning of this
introduction using these new tools. In the \textit{conventional} cases that we
have studied so far, it is known how one can arrive at (extremal) solutions
described by harmonic function from (non-extremal) solutions described by
hyperbolic sines and cosines: we will apply our new tools to a
\textit{non-conventional} (non-supersymmetric) extremal solution of the $t^3$ model not
considered in our previous works Refs.~\cite{Meessen:2012su,Galli:2012pt}. This
solution, which has been known for some time
\cite{LopesCardoso:2007ky,Gimon:2009gk,Galli:2010mg,Bossard:2012xs}, is
characterized by $H$-variables that contain anharmonic terms and its
deformation into a non-supersymmetric (finite-temperature) solution has proven
elusive \cite{Galli:2012jh}. We think that, in order to search for this
non-extremal generalization (if it exists), it is necessary to know more about
the structure of the extremal solution and we will show how the new tools can
help us to this end. 

This paper is organized as follows: in Section~\ref{sec-HFGK} we briefly
review the H-FGK formalism, providing the definitions and relations that we
will use in the rest of the article. In Section~\ref{sec-equivariantvectors} we explain
how equivariant charge vectors enter in black-hole solutions when we express
them in the $H$-variables of this formalism. In Section~\ref{sec-conventional}
we explain when the usual harmonic ansatz becomes insufficient to write the
general family of solutions associated to some attractor (expressed through an
equivariant charge vector). This insufficiency indicates the need of adding
anharmonic terms to the $H$-variables giving rise to what we have called
\textit{unconventional} black-hole solutions.  Then, in Section~\ref{sec-firstorder} we give a general form for the first-order flow equations of any static black-hole solution of these theories that applies, in particular, to the
unconventional solutions. In Sections~\ref{sec-axidilaton}
and \ref{sec-CPn} we review the supersymmetric and non-supersymmetric extremal
solutions (which are completely conventional) of two simple models, studying
their duality symmetries and their equivariant vectors. In
Section~\ref{sec-t3} we turn to the $t^{3}$ model, showing how its extremal,
non-supersymmetric solutions are non-conventional. We, then, construct and
study this unconventional family of solutions using a basis of equivariant
vectors. Our conclusions and comments on further directions of work can be
found in Section~\ref{sec-conclusions}.


\section{The H-FGK formalism for $\mathcal{N}=2$, $d=4$ supergravity}
\label{sec-HFGK}

As shown in Refs.~\cite{Mohaupt:2011aa,Meessen:2011aa}\footnote{We will follow
  the notation and conventions of Ref.~\cite{Meessen:2011aa}. More information
  about this formalism and the original FGK formalism can be found in {\em e.g.\/}
  Refs.~\cite{Ferrara:1997tw,Galli:2011fq,Galli:2012ji}. 
} 
the problem of finding static, single-center,
black-hole solutions of any ungauged $\mathcal{N}=2$, $d=4$ supergravity theory
coupled to $n$ vector multiplets can be reduced to that of finding solutions to the
effective action for the $2(n+1)$ real variables\footnote{\label{foot:1}The
  indices $M,N$ are $2(n+1)$-dimensional symplectic indices.  We use the
  symplectic metric $\left(\Omega_{MN}
  \right) \equiv \bigl(\begin{smallmatrix} 0 & \mathbbm{1}\\
    -\mathbbm{1} & 0
\end{smallmatrix} \bigr)$ and $\Omega^{MP}\Omega_{NP}= \delta^{M}{}_{N}$ to
lower and rise the symplectic indices according to the convention
\begin{equation}
  H_{M} = \Omega_{MN}H^{N}\, ,
  \hspace{1cm}  
  H^{M} = H_{N}\Omega^{NM}\, .
\end{equation}
}
$H^{M}(\tau)$

\begin{equation}
\label{eq:effectiveaction2}
-I_{\text{H-FGK}}[H] 
 =  
\int d\tau 
\left\{ 
\tfrac{1}{2}g_{MN}
\dot{H}^{M}\dot{H}^{N}
-V
\right\}\, ,
\end{equation}

\noindent
subject to the \textit{Hamiltonian constraint}

\begin{equation}
\label{eq:hamiltonianconstraint}
\tfrac{1}{2}g_{MN}
\dot{H}^{M}\dot{H}^{N}
+V
+
r^{2}_{0}
=
0\, ,
\end{equation}

\noindent
where $r_{0}$ is the \textit{non-extremality parameter}. For later
reference, we quote the equations of motion that follow from the above action,
taking into account that the metric $g_{MN}$ is not invertible 
\cite{Mohaupt:2011aa,Galli:2012ji}

\begin{equation}
\label{eq:eoms}
g_{MN}\ddot{H}^{N}
+(\partial_{N}g_{PM}-\tfrac{1}{2}\partial_{M}g_{NP})\dot{H}^{N}\dot{H}^{P}
+\partial_{M}V=0\, .
\end{equation}

The metric $g_{MN}(H)$ and the potential $V(H)$ of the H-FGK effective action
are given in terms of the \textit{Hesse potential} $\mathsf{W}(H)$ by

\begin{eqnarray}
g_{MN}(H) 
& \equiv &
\partial_{M}\partial_{N}\log{\mathsf{W}}
-2\frac{H_{M}H_{N}}{\mathsf{W}^{2}}\, ,
\\
& & \nonumber \\
\label{eq:potential}
V(H) & \equiv & 
\left\{
-\tfrac{1}{4}\partial_{M}\partial_{N}\log{\mathsf{W}}
+\frac{H_{M}H_{N}}{\mathsf{W}^{2}}
\right\}\mathcal{Q}^{M}\mathcal{Q}^{N}\, .
\end{eqnarray}

\noindent
The Hesse potential contains all the information characterizing the
$\mathcal{N}=2,d=4$ supergravity theory under consideration, and
defines it (at least in this context) just as the
canonically-normalized covariantly-holomorphic
symplectic section $(\mathcal{V}^{M}) = \bigl(\begin{smallmatrix} \mathcal{L}^{\Lambda} \\
  \mathcal{M}_{\Lambda} \end{smallmatrix} \bigr)$ does. The Hesse
potential can be derived from $\mathcal{V}^{M}$ as follows:

\begin{enumerate}
\item Introduce an auxiliary complex variable $X$ with the same K\"ahler
  weight as $\mathcal{V}^{M}$, we can define the two K\"ahler-neutral real
  symplectic vectors $\mathcal{R}^{M}$ and $\mathcal{I}^{M}$

\begin{equation}
  \label{eq:RandIdef}
\mathcal{V}^{M}/X \equiv   \mathcal{R}^{M} +i  \mathcal{I}^{M}\, .
\end{equation}

\noindent
The components of $\mathcal{R}^{M}$ can be expressed in terms of those of
$\mathcal{I}^{M}$ (solving the \textit{stabilization equations}
a.k.a.~\textit{Freudenthal duality equations} \cite{Galli:2012pt}). The
functions $\mathcal{R}^{M}(\mathcal{I})$ are characteristic of each theory,
but they are always homogeneous of first degree in the $\mathcal{I}^{M}$. 

It can be shown that 

\begin{equation}
\label{eq:Xdef}
X = \tfrac{1}{\sqrt{2}}e^{U+i\alpha}\, ,
\end{equation}

\noindent
where $e^{U}$ is the metric function (or warp factor) and $\alpha$ is a
completely arbitrary $\tau$-dependent phase which does not enter in the
Lagrangian. Different choices of $\alpha$ give different definitions of the
variables $H^{M}$ which, nevertheless, describe the same physical
variables. This freedom gives rise to a local symmetry of the H-FGK action, known as \textit{local Freudenthal duality} \cite{Galli:2012ji}, that will be
discussed later.

\item Given those functions, the \textit{Hesse potential} $\mathsf{W}(\mathcal{I})$
is just

\begin{equation}
  \mathsf{W}(\mathcal{I}) 
  \equiv 
  \mathcal{R}_{M}(\mathcal{I})\mathcal{I}^{M}\, .  
\end{equation}

\noindent
It is, by construction, homogeneous of second degree in $\mathcal{I}^{M}$.
\end{enumerate}

It is customary to relabel these variables

\begin{equation}
H^{M} \equiv \mathcal{I}^{M}\, ,
\hspace{1.5cm}
\tilde{H}^{M} \equiv \mathcal{R}^{M}\, ,
\,\,\,
\longrightarrow
\,\,\,\,
\left\{
  \begin{array}{rcl}
\mathcal{V}^{M}/X 
& = &  
\tilde{H}^{M} +iH^{M} \equiv \mathcal{H}^{M}\, .
\\
& & \\
\mathsf{W}(H) 
& = & 
\tilde{H}_{M}(H)H^{M}\, .
\\
\end{array}
\right.
\end{equation}

The relation between the tilded and untilded variables defines the
\textit{discrete Freudenthal duality} transformation of the theory
\cite{Borsten:2009zy,Ferrara:2011gv,Galli:2012ji}: $\tilde{H}^{M}$ is the
Freudenthal dual of $H^{M}$. This duality transformation turns out to be an
anti-involution, {\em i.e.\/}

\begin{equation}
\label{eq:antiinvolution}
\tilde{\tilde{H}}^{M} \equiv \tilde{H}^{M}(\tilde{H}) = -H^{M}\, ,  
\end{equation}

\noindent
and, therefore preserves the Hesse potential

\begin{equation}
\mathsf{W}(\tilde{H})  
=
\mathsf{W}(H)\, ,  
\end{equation}

\noindent
and also the full effective action
Eq.~(\ref{eq:effectiveaction2}). These discrete duality transformations are
associated to the constant shift of the phase of $X$, $\alpha\rightarrow
\alpha+ \pi$. The continuous, local, transformations $\alpha\rightarrow
\alpha+ f(\tau)$

\begin{equation}
\mathcal{H}^{\prime\, M}
=
e^{if(\tau)}\
\mathcal{H}^{M}\, ,
\end{equation}

\noindent
leave invariant the effective action (\ref{eq:effectiveaction2}) and all the physical fields \cite{Galli:2012ji}.
Since the central charge of these theories
$\mathcal{Z}(Z,Z^{*},\mathcal{Q})$ is defined in terms of the
canonically-normalized covariantly-holomorphic symplectic section
$\mathcal{V}^{M}$ by\footnote{We will often use the lighter notation
  $\mathcal{Z}(\mathcal{Q})$ or $\mathcal{Z}(B)$ if we replace the
  charge vector by another equivariant charge vector.  Sometimes these
  equivariant charge vectors are called \textit{fake charges} in which
  case $\mathcal{Z}(B)$ is referred to as \textit{fake central
    charge}.}

\begin{equation}
\label{eq:centralchargedef}
\mathcal{Z}(\mathcal{Q})\equiv \mathcal{V}_{M}\mathcal{Q}^{M}\, , 
\end{equation}

\noindent
using the definition of the $H$-variables we find that\footnote{In
  what follows, $\mathsf{W}$ with no arguments will be assumed to be
  $\mathsf{W}(H)$.}

\begin{equation}\label{eq:centralchargedefHvariables}
\mathcal{Z}(\mathcal{Q})  
= 
\frac{e^{-i\alpha}}{\sqrt{2\mathsf{W}}} \mathcal{H}_{M}\mathcal{Q}^{M}\, ,
\end{equation}

\noindent
whence under Freudenthal duality 

\begin{equation}
\mathcal{Z}^{\prime}(\mathcal{Q})  = e^{if(\tau)} \mathcal{Z}(\mathcal{Q})\, .
\end{equation}

The definition of Freudenthal dual can be extended to any symplectic
vector of a given theory\footnote{In some theories not all symplectic
  vectors have a Freudenthal dual. For instance, in the cubic models
  that we are going to study, only when the Hesse potential, evaluated
  on a particular vector, is different from zero, is the Freudenthal dual
  well defined.
} 
and, in particular, to the charge vector
$\mathcal{Q}^{M}$. We know that the black-hole potential, which is
related to the potential $V$ appearing in the H-FGK action by

\begin{equation}
\label{relationbetweenpotentials}
V_{\rm bh}= -\mathsf{W}\ V\, ,  
\end{equation}

\noindent
as a function of the variables $H^{M}$, is always extremized by the
near-horizon value $B^{M}=\beta \mathcal{Q}^{M}$ for any
proportionality constant $\beta$. Freudenthal symmetry implies that
it is also extremized at the same points in terms of the
Freudenthal-dual variables $\tilde{B}^{M}=\beta \mathcal{Q}^{M}$,
which corresponds to $B^{M} =-\beta \tilde{\mathcal{Q}}^{M}$
\cite{Ferrara:2011gv,Galli:2012ji}.  Freudenthal duality can also be
seen as a relation between black holes with identical metrics (and,
therefore, entropies) and scalar fields but different charges
\cite{Borsten:2009zy}.


\section{Explicit solutions and equivariant vectors}
\label{sec-equivariantvectors}


The main advantage of the H-FGK formalism is the linear behavior of the
variables under transformations of the electric-magnetic duality group $\mathrm{G}$ of
the theory:

\begin{equation}
H^{M\, \prime} = S^{M}{}_{N} H^{N}\, ,  
\hspace{1cm}
(S^{M}{}_{N}) \in \mathrm{G} \subset \mathrm{Sp}(2n+2;\mathbb{R})\, .
\end{equation}

\noindent
This linear behavior can dramatically simplify the construction of explicit solutions
to theories with a non-trivial duality group as it implies that any
solution must be of the form 

\begin{equation}
\label{eq:HMform}
H^{M}(\tau) = c^{\sigma}(\tau)\ U_{\sigma}^{M}\, ,  
\end{equation}

\noindent 
where the functions $c^{\sigma}(\tau)$ are duality invariant; the
symplectic vectors $U_{\sigma}^{M}$ are constant vectors that may depend on
the physical parameters of the theory (mass $M$, electric and magnetic charges
$\mathcal{Q}^{M}$ and asymptotic values of the scalars $Z^{i}_{\infty}$) and
must be \textit{equivariant} w.r.t.~the duality group, \textit{i.e.}

\begin{equation}
U_{\sigma}^{M} (M,Z_{\infty}^{\prime},Z^{*\, \prime}_{\infty},\mathcal{Q}^{\prime}) 
=   
S^{M}{}_{N} U_{\sigma}^{N}(M,Z_{\infty},Z^{*}_{\infty},\mathcal{Q})\, ,
\end{equation}

\noindent
with

\begin{equation}
Z^{i\, \prime}\equiv F_{S}^{i}(Z)\, , 
\hspace{1cm}
\mathcal{Q}^{M\, \prime} = S^{M}{}_{N} \mathcal{Q}^{N}\, ,
\end{equation}

\noindent
where $F_{S}^{i}(Z)$ is the non-linear realization of the duality
transformation $S^{M}{}_{N}$ on the complex scalars.

In some cases, the number of equivariant vectors of the theory can be
greater than\footnote{If it is greater, we can eliminate some from the ansatz,
  since they will be linearly dependent on the rest.
} 
or equal to the
number of variables $H^{M}$. In that case, one does not win much by using the
above ansatz.  In other cases, however, the number can be much smaller and
we will be left with a small number of invariant functions to be determined.

In the near-horizon limit of extremal black-hole solutions, the value of the
variables $H^{M}$ will be dominated by one equivariant vector that we denote
by $B^{M}$ and that can be defined, in our conventions, by\footnote{Observe
  that this definition is completely general: given the behavior of the
  3-dimensional transverse metric in the near-horizon limit as a function of
  $\tau$ and the degree of homogeneity of $e^{-2U}=\mathsf{W}(H)$ as a
  function of the $H$-variables, in regular black-hole solutions the functions
  $H^{M}(\tau)$ are dominated by these constant vectors in the near-horizon
  limit.}

\begin{equation}
  B^{M} \equiv \lim_{\tau \rightarrow -\infty} -\frac{\sqrt{2} H^{M}}{\tau}\, .   
\end{equation}

\noindent
The values of the scalars on the horizon, $Z^{i}_{\rm h}$, are completely
determined by this equivariant vector upon use of the general expression of
the scalars as functions of the variables $H^{M}$ \cite{Meessen:2011aa}

\begin{equation}
Z^{i}(H)  = \frac{\tilde{H}^{i}+iH^{i}}{\tilde{H}^{0}+iH^{0}}\, , 
\hspace{.5cm}
\Rightarrow
\hspace{.5cm}
Z^{i}_{\rm h}=Z^{i}(B)\, ,
\end{equation}

\noindent
and also extremize the black-hole potential $V_{\rm bh}(H,\mathcal{Q})$ as a
function of the variables $H^{M}$:

\begin{equation}
\left. \partial_{M} V_{\rm bh}(H,\mathcal{Q})  \right|_{H=B} =0\, .
\end{equation}

The vectors $B^{M}$, which in this context can be called attractors,
can also be written in the form

\begin{equation}
B^{M} = b^{\sigma} U_{\sigma}^{M}\, ,  
\end{equation}

\noindent
where the $b^{\sigma}$ are duality-invariant constants such that the
products $bU^{M}$ have the same dimensions as electric and magnetic
charges. 

Clearly these vector attractors must contain more information than the values
of the scalars on the horizon $Z^{i}_{\rm h}$ (the standard
attractors).  On the other hand, when the model has a high degree of
symmetry the requirement of equivariance imposes strong constraints on
the possibilities and it simplifies the task of finding the attractors
$B^{M}$.

A similar discussion can be made for the values of the variables
$H^{M}$ at spatial infinity, which in the employed coordinate system lies at $\tau =0$. 

The amount of simplification introduced by the above observation that
the variables $H^{M}$ must always be of the form Eq.~(\ref{eq:HMform})
depends on our ability to find a sufficient number of equivariant
vectors; the Freudenthal dual of the charge vector
$\tilde{\mathcal{Q}}^{M}$ is, by construction, a prime example of
equivariant vector, but there are other systematic ways of finding
them. Let us consider, first, equivariant vectors that only depend on
the charges. They can be seen as an endomorphism of the
$(2n+2)$-dimensional vector space of charges and their equivariance is
equivalent to the fact that these endomorphisms commute with the duality
transformations (which are also endomorphisms of charge space). Thus,
linear (not necessarily symplectic) transformations that commute with $\mathrm{G}$
provide a second example of equivariant vectors.

To study non-linear cases, let us expand an equivariant vector and the
duality transformations around the identity

\begin{equation}
U^{M}_{\sigma}(\mathcal{Q})
\sim 
\mathcal{Q}^{M} +\xi^{M}(\mathcal{Q})\, ,
\hspace{1cm}
(S\mathcal{Q})^{M} 
\sim  
\mathcal{Q}^{M} +\alpha^{A}\eta_{A}{}^{M}(\mathcal{Q})\, ,
\end{equation}

\noindent
where $S \in \mathrm{G} \subset \mathrm{Sp}(2n+2;\mathbb{R})$ and, therefore,

\begin{equation}
\eta_{A}{}^{M}(\mathcal{Q}) 
\ =\ 
(T_{A})^{M}{}_{N}\ \mathcal{Q}^{N}\, , 
\end{equation}

\noindent
where $T_{A}\in \mathrm{Sp}(2n+2;\mathbb{R})$ are the generators of the
duality group; the condition of equivariance is equivalent to
requiring that the Lie brackets of these two kinds of generators
vanish\footnote{Obviously, also $\xi$ must be an equivariant vector, whence
  we can replace $\xi$ by $U$ in what follows for the
  purpose of writing an equation characterizing equivariant vectors.}

\begin{equation}
\label{eq:equivariantvectorequation}
[U,\eta_{A}] =0\, ,
\,\,\,\,   
\Rightarrow
\,\,\,\,   
(T_{A})^{M}{}_{N} \mathcal{Q}^{N} \partial_{M}U^{P}
=
(T_{A})^{P}{}_{R}U^{R}\, ,
\,\,\,\,\,
\mathrm{where}
\,\,\,\,\,
\partial_{M}U^{P} \equiv \frac{\partial U^{P}}{\partial\mathcal{Q}^{M}}\, .
\end{equation}

On taking the derivative with respect to $\mathcal{Q}^{P}$ of both sides
of this equation we find the integrability condition
 
\begin{equation}
(T_{A})^{M}{}_{N} \mathcal{Q}^{N} \partial_{M} \mathsf{P}
=
0\, ,  
\hspace{1cm}
\mathsf{P} \equiv \partial_{M} U^{M}= \Omega^{MN}\partial_{M} U_{N}\, .
\end{equation}

\noindent
which implies that $\mathsf{P}$ is an invariant function of the charges. Thus,
equivariant vectors are associated to invariants by the above
equation.  The simplest invariant is just $\mathsf{P}=0$ and equivariant
vectors such that $\partial_{[M}U_{N]}=0$ are associated to it;
clearly there may be more possibilities as locally they must be of the
form $U_{M} = \partial_{M} h$ for some non-vanishing invariant $h$
(possibly up to additive numerical constants) and one can check that
the equivariance condition is automatically satisfied. For instance,
if we take $h = \mathsf{W}/2$, then $U_{M}=\tilde{\mathcal{Q}}_{M}$.

For equivariant vectors that depend (non-holomorphically) on the
moduli $Z^{i}_{\infty}$, the equivariance condition takes the form

\begin{equation}
  (T_{A})^{M}{}_{N} \mathcal{Q}^{N} \partial_{M}U^{P}
  +
  k_{A}{}^{i}\partial_{i}U^{P}
  +
  k_{A}{}^{*\, i^{*}}\partial_{i^{*}}U^{P}
  =
  (T_{A})^{P}{}_{R}U^{R}\, ,  
\end{equation}

\noindent
where $K_{A}\equiv k_{A}{}^{i}(Z)\partial_{i} +\mathrm{c.c.}$ are the
Killing vectors that generate the action of the duality group $\mathrm{G}$ on the
scalar manifold preserving the holomorphic and K\"ahler structures.
Again, $\mathsf{P}\equiv \partial_{M}U^{M}$ must be an invariant and a
particularly simple case is $\mathsf{P}=0$ and $U_{M}=\partial_{M} h$ where,
now, $h$ is required to be invariant only up to additive functions of
the moduli. A recurring example is

\begin{equation}
h\ =\ \log\ \left( \mathcal{Z}(\mathcal{Q})\right)\, ,
\end{equation}

\noindent
where $\mathcal{Z}(\mathcal{Q})$ is the central charge defined in
Eq.~(\ref{eq:centralchargedef}). The associated (complex) equivariant
vector is

\begin{equation}
\label{eq:genericmodulidependentequivariantvector}
U_{M}
=
\frac{\partial\, h}{\partial \mathcal{Q}^{M}} 
=
\frac{\mathcal{V}_{M}}{\mathcal{Z}(\mathcal{Q})}\, . 
\end{equation}

\noindent
The real and imaginary parts provide two real moduli-dependent
equivariant vectors.  It should be obvious that one can use, instead
of the central charge any fake central charge, but the result may not
be a new equivariant vector.

The Lie bracket of two equivariant vectors is also an equivariant
vector, so that the equivariant vectors form a Lie algebra that commutes with that
of the duality group $\mathrm{G}$.

Finally, in the cases that we are going to study, we will show how one
can construct equivariant vectors by using other methods like
solution-generating techniques.


\section{Conventional and unconventional solutions}
\label{sec-conventional}


As explained in Ref.~\cite{Meessen:2011aa}, contracting the equations of
motion derived from the H-FGK action Eq.~(\ref{eq:effectiveaction2}) with
$H^{M}$ and using the homogeneity properties of the different terms and the
Hamiltonian constraint Eq.~(\ref{eq:hamiltonianconstraint}) one finds, in the
extremal case $r_{0}=0$\footnote{In this discussion we will only consider the
  extremal case because in the rest of the paper we are going to restrict
  ourselves to it.}, the equation

\begin{equation}
\label{eq:Urewriten}
\mathsf{W}\tilde{H}_{M}\, \ddot{H}^{M}
+
(\dot{H}^{M}H_{M})^{2}  
=
0\, .
\end{equation}

In what we are going to call from now on \textit{conventional} extremal
solutions (supersymmetric or not) the variables $H^{M}(\tau)$ are harmonic
functions, {\em i.e.\/} they satisfy $\ddot{H}^{M}=0$. The above equation implies that they
also satisfy the constraint\footnote{The converse is not always true: the
  above constraint can be satisfied for extremal black-hole solutions which
  are not given by harmonic $H^{M}$s and that we will call
  \textit{unconventional}.  }

\begin{equation}
\label{eq:nonut}
\dot{H}^{M}H_{M} = 0\, . 
\end{equation}

\noindent
Conventional extremal solutions have been intensively studied in
Ref.~\cite{Galli:2012pt}. However, how general are these solutions? Can all
the extremal black-hole solutions be written in a conventional form? (The
answer in the supersymmetric case is yes.) If not, what are the limitations
and how can they be overcome as to obtain the most general extremal black-hole
solutions that depend on the maximal number of independent physical parameters?

To investigate these issues, it is convenient to review in detail the
construction of conventional extremal black-hole solutions:
extremal black-holes are associated to values of the scalar fields $Z^{i}_{\rm
  h}$ (attractors) that extremize the black-hole potential
\cite{Ferrara:1997tw}. As explained in the previous section, in the H-FGK
formulation attractors appear as symplectic vectors $B^{M}$ that extremize the
black-hole potential when written in terms of the $H$-variables. These attractors
$B^{M}$ are defined up to normalization because the black-hole potential is
invariant under rescalings of the $H^{M}$s and also up to global Freudenthal
rotations. Furthermore, as functions of the charges and moduli, the attractors
$B^{M}$ are equivariant under duality transformations. A family of
extremal black holes closed under duality will be associated to a given
equivariant vector expressed as a set of functions of the charge components
and moduli $B^{M}(\mathcal{Q},Z_{\infty},Z^{*}_{\infty})$. We are going to
focus on moduli-independent attractors, {\em i.e.\/} the so-called \textit{true attractors}.


The attractor $B^{M}$ determines the near-horizon form of the solution. We can
always construct a solution describing the AdS$^{2}\times$S$_{2}$ solution
that describes the near-horizon geometry by choosing the appropriate
normalization of $B^{M}$: indeed, one can check that the harmonic functions

\begin{equation}
\label{eq:NHsolution}
H^{M} = -\tfrac{1}{\sqrt{2}}B^{M}\tau\, ,  
\end{equation}

\noindent
always satisfy the equations of motion as long as the condition

\begin{equation}
\label{eq:conditionNHsolution}
V_{\rm bh}(B,\mathcal{Q}) = -\tfrac{1}{2}\mathsf{W}(B)\, ,  
\end{equation}

\noindent
determining the normalization of $B^{M}$ is met.

To construct a solution with the same near-horizon behavior and with an
asymptotically-flat region we must add to the $H^{M}$ above a constant vector
$A^{M}$. The condition Eq.~(\ref{eq:nonut}) and the normalization of the
metric at infinity become two constraints for $A^{M}$

\begin{equation}
\label{eq:genericconstraintsonAM}
B^{M}A_{M}=0\, , 
\hspace{1cm}  
\mathsf{W}(A)=1\, ,
\end{equation}

\noindent
that leave $2n$ real constants, which is just the right amount to
describe the asymptotic values of the $n$ complex scalars
$Z^{i}_{\infty}$. Only if we cannot add a vector $A^{M}$ satisfying
these two constraints, then the most general solution associated to
the attractor $B^{M}$ cannot be conventional and we will have to add
anharmonic terms to the $H^{M}$.

We can reformulate this question as follows: if we add to the $H^{M}$ in
Eq.~(\ref{eq:NHsolution}) an infinitesimal vector $\varepsilon^{M}$
satisfying $B^{M}\varepsilon_{M}=0$, do we get another solution to the
Hamiltonian constraint Eq.~(\ref{eq:hamiltonianconstraint}) and equations of
motion Eq.~(\ref{eq:eoms})? To first order in $\varepsilon^{M}$, the
Hamiltonian constraint will be solved by the perturbed solution

\begin{equation}
\label{eq:perturbedNHsolution}
H^{\prime\, M}  = H^{M} +\varepsilon^{M}\, ,
\hspace{1cm}
H^{M} = -\tfrac{1}{\sqrt{2}}B^{M}\tau\, ,
\hspace{1cm}
B^{M}\varepsilon_{M}=0\, ,
\end{equation}

\noindent
if

\begin{equation}
\varepsilon^{M}
\left\{
\tfrac{1}{2}\partial_{M}g_{NP} \dot{H}^{N}\dot{H}^{P}
+\partial_{M}V(H,\mathcal{Q})
\right\}
=
0\, .   
\end{equation}

\noindent
Evaluating this equation at the near-horizon solution $H^{M}$, using $V_{\rm
  bh}(H,\mathcal{Q})= -\mathsf{W}(B)V(H,\mathcal{Q})$, the homogeneity
properties of the different terms, the fact that $\partial_{M}V_{\rm
  bh}(B,\mathcal{Q})=0$ and the condition (\ref{eq:conditionNHsolution}),
we arrive at

\begin{equation}
\varepsilon^{M}
\left\{
\tfrac{1}{4}B^{N}B^{P}\partial_{M}\partial_{N}\partial_{P}\log{\mathsf{W}}(B)
-\tfrac{1}{2}\partial_{M}\log{\mathsf{W}}(B)
\right\}
=
0\, ,
\end{equation}

\noindent
which is an equation in the variables $B^{M}$ (including the partial
$\partial_{M}$ derivatives, which should be understood as partial derivatives
with respect to $B^{M}$) and is identically satisfied on account of the scale
invariance of $\log{\mathsf{W}(B)}$.

The analogous condition on the equations of motion, Eqs.~(\ref{eq:eoms}), reads

\begin{equation}
\varepsilon^{M}
\left\{
\partial_{M}g_{NP}\ddot{H}^{P}
+
\partial_{M}
(\partial_{P}g_{QN}-\tfrac{1}{2}\partial_{N}g_{PQ})
\dot{H}^{P}\dot{H}^{Q}
+\partial_{M}\partial_{N}V(H,\mathcal{Q})
\right\}
=
0\, ,  
\end{equation}

\noindent
and, after evaluation on the near-horizon solution we get a homogenous
equation that, again, can be read as an equation on the variables
$B^{M}$. Using the same properties we used with the Hamiltonian constraint
plus $B^{M}\varepsilon_{M}=0$ we get a non-trivial equation for
$\varepsilon^{M}$

\begin{equation}
\label{eq:conditioninfinitesimalextension}
\mathfrak{M}_{MN}
\varepsilon^{N}
=
0\, ,
\,\,\,\,\,
\mathrm{with}
\,\,\,\,\,
\mathfrak{M}_{MN} 
\equiv
\mathsf{W}(B)\partial_{M}\partial_{N}\log{\mathsf{W}}(B)
+2\frac{\tilde{B}_{M}\tilde{B}_{N}}{\mathsf{W}(B)}
-\partial_{M}\partial_{N}V_{\rm bh}(B,\mathcal{Q})\, .
\end{equation}

\noindent
We are interested in the number of independent solutions to this
equation that satisfy the constraint $B^{M}\varepsilon_{M}=0$,
\textit{i.e.}~in the rank of $\mathfrak{M}_{MN}$. The rank should be
at most $1$ as this implies a single linear constraint on the
components of $\varepsilon^{M}$, which should be equivalent to
$B^{M}\varepsilon_{M}=0$. If the rank of $\mathfrak{M}_{MN}$ happens
to be bigger than 1, then there are not enough unconstrained
components of $\varepsilon^{M}$ for the family of solutions to have
arbitrary values of the moduli and the most general solution based on
the chosen attractor, must necessarily contain anharmonic
terms. 

For cubic models, the need of anharmonic ans\"atze to construct the
most general, generating, non-supersymmetric, extremal, black-hole
solution of \cite{LopesCardoso:2007ky} and \cite{Gimon:2009gk} was
first observed in \cite{Galli:2010mg} and later confirmed in
\cite{Bossard:2012xs} and \cite{Galli:2012jh}. In the next sections we
will see how the obstruction to the fully harmonic ansatz arises in
the particular case of the $t^{3}$ model.  For the non-extremal case
of these theories, the situation is still unclear \cite{Galli:2012jh}.


\section{The general first-order flow equations}
\label{sec-firstorder}


The central charge of an $\mathcal{N}=2,d=4$ supergravity theory is
defined by Eq.~(\ref{eq:centralchargedef}) and, in terms of the
$H$-variables it takes the form of
Eq.~(\ref{eq:centralchargedefHvariables}) which we copy here for
convenience

\begin{equation}
  \mathcal{Z}(\mathcal{Q})  
  = 
  \frac{e^{-i\alpha}}{\sqrt{2\mathsf{W}}} 
(\tilde{H}_{M}+iH_{M})\mathcal{Q}^{M}\,.
\end{equation}

Let us consider a generalization of the central charge, denoted by
$\mathcal{Z}(\phi,\sqrt{2}\mathfrak{D}H)$, in which we replace the
second argument (the charge vector) by the Freudenthal-covariant
derivative of $H^{M}$ introduced in Ref.~\cite{Galli:2012ji}, \textit{i.e.}

\begin{equation}
\mathfrak{D}H^{M} \equiv \dot{H}^{M} +A\tilde{H}^{M}\, ,
\hspace{1cm}
A \equiv \frac{\dot{H}^{N}H_{N}}{\mathsf{W}}\, . 
\end{equation}

\noindent
Since $H_{M}\mathfrak{D}H^{M}=0$ and $\tilde{H}_{M}\tilde{H}^{M}=0$
identically, we immediately find that

\begin{equation}
|\mathcal{Z}(\phi,\sqrt{2}\mathfrak{D}H)| 
= 
\pm \frac{\tilde{H}_{M}\dot{H}^{M}}{\sqrt{\mathsf{W}}}
=  
\pm \frac{\partial_{M}\mathsf{W}\, \dot{H}^{M}}{2\sqrt{\mathsf{W}}}
=
\pm\frac{d \sqrt{\mathsf{W}}}{d\tau} 
=
\pm \frac{d e^{-U}}{d\tau}\, , 
\end{equation}

\noindent
which is the first-order equation for the metric
function\footnote{This equation reduces to Eq.~(5.9) of
  Ref.~\cite{Galli:2010mg} in the extremal limit. Observe that the
  Freudenthal-covariant derivative corresponds to Eq.~(5.6) of the
  same reference.}. Observe that $H_{M}\mathfrak{D}H^{M}=0$ implies
that the phase of $\mathcal{Z}(\phi,\sqrt{2}\mathfrak{D}H)$ is equal
to the phase of $\pm X$. The sign must be chosen so as to make
$\pm\tilde{H}_{M}\dot{H}^{M}>0$ and, since the mass of the solution
corresponding to $e^{-2U}=\mathsf{W}(H)$ is given by

\begin{equation}
\label{eq:Mdef}
M
= 
-\tfrac{1}{2}\left. \frac{de^{-2U}}{ d\tau}\right|_{\tau=0}
=  
-\tfrac{1}{2}\left.\dot{\mathsf{W}}\right|_{\tau=0}
=  
-\left. \tilde{H}_{M}\dot{H}^{M}\right|_{\tau=0}\, ,
\end{equation}

\noindent
we find that for regular solutions (with positive mass) we must choose
the lower sign:

\begin{equation}
\label{eq:firstorderU}
\frac{d e^{-U}}{d\tau} 
= 
-|\mathcal{Z}(\phi,\sqrt{2}\mathfrak{D}H)|\, .  
\end{equation}

From Eq.~(2.8) of Ref.~\cite{Ortin:2011vm} we have that 

\begin{equation}
\frac{dZ^{i}}{d\tau}
=  
-2X\mathcal{G}^{ij^{*}} \mathcal{D}_{j^{*}}\mathcal{V}^{*}_{M}\dot{H}^{M}\, .
\end{equation}

\noindent
We can rewrite $\dot{H}^{M}$ as

\begin{equation}
\dot{H}^{M} 
= 
\mathfrak{D}H^{M}-A\tilde{H}^{M} 
= 
\mathfrak{D}H^{M}
-A\left(\frac{\mathcal{V}^{M}}{2X}+\mathrm{c.c.} \right)\, ,
\end{equation}

\noindent
and plug it into the previous equation to get 

\begin{equation}
\label{eq:firstorderZi}
\begin{array}{rcl}
{\displaystyle\frac{dZ^{i}}{d\tau}}
& = &
-2X\mathcal{G}^{ij^{*}} 
\mathcal{D}_{j^{*}}\mathcal{Z}^{*}(\phi,\mathfrak{D}H)
=
4X e^{-i\alpha}\mathcal{G}^{ij^{*}} 
\partial_{j^{*}}|\mathcal{Z}^{*}(\phi,\mathfrak{D}H)|
\\
& & \\
& = &
2e^{U}\mathcal{G}^{ij^{*}} 
\partial_{j^{*}}|\mathcal{Z}^{*}(\phi,\sqrt{2}\mathfrak{D}H)|
\, ,
\end{array}
\end{equation}

\noindent
where we have used Eq.~(\ref{eq:Xdef}) and the equality of the phases of
$-X$ and $|\mathcal{Z}(\phi,\sqrt{2}\mathfrak{D}H)|$. This is the
second first-order equation\footnote{Again, this equation reduces to
  Eq.~(5.10) of Ref.~\cite{Galli:2010mg} in the extremal limit.}.

Some remarks are in order:

\begin{enumerate}

\item In these derivations we have assumed neither extremality or
  non-extremality of the solutions nor any explicit form of the
  variables $H^{M}$ (harmonic or hyperbolic)\footnote{Actually, we
    have written \textit{solutions} but we have not used at any moment
    the fact that the $H^{M}$ solve the equations of motion. The
    first-order equations that we have derived are, therefore, valid
    for any configuration of the variables $H^{M}$, although their use
    is essentially limited to solutions.}. Furthermore, we have not
  assumed the Freudenthal gauge-fixing condition
  $\dot{H}^{N}H_{N}=0$. Only the properties of Special Geometry
  encoded in the H-FGK formalism have been used. Therefore, the
  first-order Eqs.~(\ref{eq:firstorderU}) and (\ref{eq:firstorderZi})
  apply to any static black-hole solution of ungauged
  $\mathcal{N}=2,d=4$ supergravity coupled to vector multiplets.

\item These first-order equations reduce to those found in the
  literature starting from Ref.~\cite{Ferrara:1997tw} in the
  extremal/harmonic (\textit{i.e.}~$A=\dot{H}^{N}H_{N}=0$) cases: if
  $H^{M} = A^{M}-\tfrac{1}{\sqrt{2}}B^{M}\tau$ for some constant
  symplectic vectors $A^{M}$ (which encode the values of the scalars
  at spatial infinity) and the attractor $B^{M}$, then

\begin{equation}
|\mathcal{Z}(\phi,\sqrt{2}\mathfrak{D}H)|  
=
|\mathcal{Z}(\phi,B)|\, , 
\end{equation}

\noindent
which is known as \textit{fake central charge} when $B^{M}\neq
\mathcal{Q}^{M}$ and coincides with the central charge in the
supersymmetric case $B^{M}= \mathcal{Q}^{M}$.

\item In the general (non-supersymmetric) case $\mathfrak{D}H$ will be
  $\tau$-dependent and its near-horizon ($\tau\rightarrow -\infty$)
  and spatial infinity ($\tau\rightarrow 0^{-}$) limits, will not
  necessarily be equal: in the near-horizon limit $\lim_{\tau
    \rightarrow -\infty}\mathfrak{D}H^{M} \equiv
  -\tfrac{1}{\sqrt{2}}B^{M}$ and in the spacelike infinity limit
  $\lim_{\tau \rightarrow 0^{-}}\mathfrak{D}H^{M} \equiv
  -\tfrac{1}{\sqrt{2}}E^{M}$ and, generically, $B^{M}\neq E^{M}$.

\begin{eqnarray}
M 
& = & 
-\lim_{\tau\rightarrow 0^{-}} \frac{d e^{-U}}{d\tau}
= 
 |\mathcal{Z}(\phi_{\infty},E)|\, ,\\
& & \nonumber \\
S 
& = & 
\pi \left[\lim_{\tau\rightarrow -\infty} \frac{d e^{-U}}{d\tau}\right]^{2}
=
\pi |\mathcal{Z}(\phi_{\rm h},B)|^{2}\, ,  
\end{eqnarray}

\noindent
where $\phi_{\infty}$ and $\phi_{\rm h}$ are the values of the scalars
at spatial infinity and on the horizon, respectively. Different fake
central charges $\mathcal{Z}(\phi,E)$ and $\mathcal{Z}(\phi,B)$ drive
the metric function in the spatial-infinity and near-horizon regions,
respectively. This behavior is present in the non-supersymmetric
extremal solutions of the cubic models studied in
Refs.~\cite{LopesCardoso:2007ky,Gimon:2007mh,Bena:2009ev,Galli:2010mg,Bossard:2012xs}
which have anharmonic $H^{M}$s\footnote{The $H^{M}$s of those
  solutions do not satisfy the constraint $\dot{H}^{M}H_{M}=0$. A
  change of Freudenthal gauge can bring the solutions to the
  $\dot{H}^{M}H_{M}=0$ gauge but cannot make the $H^{M}$ harmonic
  \cite{Galli:2012ji}.}.

\item In Ref.~\cite{Ceresole:2007wx} and subsequent literature the
  first-order flow equations were given in terms of superpotential
  functions $W(\phi,B)$ which depend only on a constant fake charge
  vector $B^{M}$ and which has a structure similar, but not identical,
  to the central charge. Those first-order equations must be
  completely equivalent to
  Eqs.~(\ref{eq:firstorderU},\ref{eq:firstorderZi}), because the same
  variables, for the same solution, cannot obey two different sets of
  first-order equations. We do not know how to prove this equivalence
  in general, and it will have to be checked case by case.

\end{enumerate}


\section{The axidilaton model}
\label{sec-axidilaton}


The axidilaton model is defined by the prepotential

\begin{equation}
\mathcal{F}=-i\mathcal{X}^{0}\mathcal{X}^{1}\, ,  
\end{equation}

\noindent
and has only one complex scalar that we will denote by $\lambda$ that
is given by

\begin{equation}
\lambda\equiv i\mathcal{X}^{1}/\mathcal{X}^{0}\, .
\end{equation}

\noindent
In terms of $\lambda$ and in the $\mathcal{X}^{0}=i/2$ gauge, the K\"ahler
potential and metric are

\begin{equation}
\label{eq:kahlermetricandpotentialaxidilaton}
\mathcal{K}=-\ln{\Im\mathfrak{m}\lambda}\, ,  
\hspace{1cm}
\mathcal{G}_{\lambda\lambda^{*}} = (2\Im\mathfrak{m}\lambda)^{-2}\, ,
\end{equation}

\noindent
and therefore $\lambda$, which must take values in the upper half complex
plane, parametrizes the coset space $\mathrm{Sl}(2;\mathbb{R})/\mathrm{SO}(2)$. 

The canonically-normalized covariantly-holomorphic symplectic section
$\mathcal{V}$ is, in the gauge in which the K\"ahler potential is
given by Eq.~(\ref{eq:kahlermetricandpotentialaxidilaton}),

\begin{equation}
\label{eq:sectionaxidilaton}
\mathcal{V}=\frac{1}{2(\Im {\rm m}\lambda)^{1/2}}
\left( 
  \begin{array}{c}
i \\ \lambda \\ -i\lambda \\ 1 \\
  \end{array}
\right)\, ,  
\end{equation}

\noindent
and the central charge and its holomorphic covariant derivative are

\begin{equation}
  \begin{array}{rcl}
\mathcal{Z}(\mathcal{Q})
& = &  
{\displaystyle\frac{1}{2\sqrt{\Im {\rm m}\lambda}}}
\left[\, (p^{1} -i q_{0})-(q_{1} +i p^{0})\lambda\, \right]\, ,
\\
& & \\
\mathcal{D}_{\lambda}\mathcal{Z} 
& = & 
{\displaystyle\frac{i}{4(\Im {\rm m}\lambda)^{3/2}}}
\left[\, (p^{1} -i q_{0})-(q_{1} +i p^{0})\lambda^{*}\, \right]\, .
\end{array}
\end{equation}

It is useful to define the fake charge and associated fake central charge 

\begin{equation}
\label{eq:fakecentralcharge1}
\mathcal{P}
\equiv
\left(
  \begin{array}{c}
   p^{0} \\ -p^{1} \\ q_{0} \\ -q_{1} \\ 
  \end{array}
\right)\, ,
\hspace{1cm}
\mathcal{Z}(\mathcal{P})
\equiv 
{\displaystyle\frac{1}{2\sqrt{\Im {\rm m}\lambda}}}
\left[\,  (-p^{1} -i q_{0})-(-q_{1} +i p^{0})\lambda\, \right]\, ,  
\end{equation}

\noindent
in terms of which 

\begin{equation}
\mathcal{G}^{ij^{*}}\mathcal{D}_{i}\mathcal{Z}\mathcal{D}_{j^{*}}\mathcal{Z}^{*} 
=
|\mathcal{Z}(\mathcal{P})|^{2}\, ,  
\end{equation}

\noindent
so that the 
black-hole potential takes the simple form 

\begin{equation}
\label{eq:V=Z+Z}
-V_{\rm bh}
=  
|\mathcal{Z}(\mathcal{Q})|^{2}+|\mathcal{Z}(\mathcal{P})|^{2}\, .
\end{equation}

The black-hole solutions of this model have been exhaustively studied
in
Refs.~\cite{kn:axidilaton,Kallosh:1993yg,LozanoTellechea:1999my,Galli:2011fq}.
Our goal here is to illustrate the general results and methods
described in the previous sections using this well-known model. First,
let us recall what are the symmetries of this model in its original
formulation.


\subsection{The global symmetries of the axidilaton model}
\label{sec-globalsymaxidil}


The full axidilaton model (and not just the axidilaton kinetic term)
is invariant under global $\mathrm{Sl}(2;\mathbb{R})$ transformations. Let us
start by describing the action of this group on the axidilaton field:
parametrize a generic element of $\mathrm{Sl}(2;\mathbb{R})$ as

\begin{equation}
\label{eq:SL2Rmatrix}
\Lambda
\equiv
\left(
  \begin{array}{cc}
a & b \\ c  & d  \\  
  \end{array}
\right)\,, \hspace{1cm}\mbox{with}\;\;\; ad-bd =1 \, ,
\end{equation}

\noindent
then the axidilaton transforms as 

\begin{equation}
\label{eq:finitesl2rtransformations}
\lambda^{\prime} = \frac{a\lambda +b}{c\lambda +d}\, .
\end{equation}

\noindent
The scalar manifold metric admits 3 holomorphic Killing vectors which can be
taken to be

\begin{equation}
\label{eq:taukillingvectors}
K_{1} = \lambda\partial_{\lambda} +\mathrm{c.c.}\, ,
\hspace{.7cm}  
K_{2} = \tfrac{1}{2}(1-\lambda^{2})\partial_{\lambda} +\mathrm{c.c.}\, ,
\hspace{.7cm}  
K_{3} = \tfrac{1}{2}(1+\lambda^{2})\partial_{\lambda} +\mathrm{c.c.}\, ,
\end{equation}

\noindent
and satisfy the commutation relations of the Lie algebra $\mathfrak{sl}(2;\mathbb{R})$

\begin{equation}
[K_{m},K_{n}] = \epsilon_{mnq}\eta^{qp} K_{p}\ ,
\,\,\,\,\,\,\,
\Rightarrow f_{mn}{}^{p} = -\epsilon_{mnq}\eta^{qp},\,\,\,\,\,
(m,n,\ldots =1,2,3)\ ,  
\end{equation}

\noindent
where $\epsilon_{123}=+1$ and $\eta = \mathrm{diag}(++-)$; 
$\eta$ is proportional to the Killing metric of $\mathfrak{so}(1,2)\simeq\mathfrak{sl}(2;\mathbb{R})\simeq\mathfrak{sp}(2;\mathbb{R})$.
The infinitesimal $\mathrm{Sl}(2;\mathbb{R})$
transformations of $\lambda$ can be written using these Killing
vectors as

\begin{equation}
  \delta_{\alpha}\lambda = \alpha^{m}k_{m}{}^{\lambda} = 
  \tfrac{1}{2}(\alpha^{2}+\alpha^{3}) +\alpha^{1}\lambda  
  -\tfrac{1}{2}(\alpha^{2}-\alpha^{3})\lambda^{2}\, .
\end{equation}

\noindent
The infinitesimal linear transformations associated to the above choice of
Killing vectors is, in terms of the Pauli matrices

\begin{equation}
\left(
  \begin{array}{cc}
a~ & ~b \\ c~  & ~d  \\  
  \end{array}
\right)  
\sim 
\mathbbm{1}_{2\times 2}
+ \alpha^{m}T_{m}\, ,
\hspace{1cm}
T_{1}= -\tfrac{1}{2}\sigma^{3}\, ,\,\,\,
T_{2}= -\tfrac{1}{2}\sigma^{1}\, ,\,\,\,
T_{3}= \tfrac{i}{2}\sigma^{2}\, ,
\end{equation}

\noindent
and satisfy the Lie algebra

\begin{equation}
[T_{m},T_{n}] = -\epsilon_{mnq}\eta^{qp} T_{p}\, . 
\end{equation}

The action of the finite $\mathrm{Sl}(2;\mathbb{R})$ transformations on the
K\"ahler potential and on the canonical covariantly-holomorphic
symplectic section $\mathcal{V}$ given in
Eq.~(\ref{eq:sectionaxidilaton}) is 

\begin{eqnarray}
\label{eq:Kduality}
\mathcal{K}^{\prime}(\lambda) 
& \equiv  &
\mathcal{K}(\lambda^{\prime}(\lambda))
=
\mathcal{K}(\lambda) + 2\Re\mathfrak{e} f(\lambda)\, ,
\\
& & \nonumber \\
\label{eq:Vduality}
\mathcal{V}^{\prime\, M}(\lambda) 
& \equiv &  
\mathcal{V}^{M}(\lambda^{\prime}(\lambda))
=
e^{ -i\Im\mathfrak{m} f(\lambda)} \
S^{M}{}_{N}
\mathcal{V}^{N}\, , 
\end{eqnarray}

\noindent
where the holomorphic function $f(\lambda)$ of the K\"ahler transformation and
the symplectic rotation $S^{M}{}_{N}$ are given by

\begin{eqnarray}
f(\lambda) 
& = &
\ln{(c\lambda+d)}\, ,
\\
& & \nonumber \\
\label{eq:sl2embeddedinsp4}
(S^{M}{}_{N})
& = &
\left(
  \begin{array}{cccc}
   d &   & -c &   \\
     & a &    & b \\
  -b &   & a  &   \\
     & c &    & d \\ 
  \end{array}
\right)\, .
\end{eqnarray}

In this 4-dimensional representation the infinitesimal generators $T_{m}$ are given by

\begin{equation}
\label{eq:symplecticTi}
(T_{1}{}^{M}{}_{N})
= 
-\tfrac{1}{2} 
\left( 
  \begin{array}{cc}
  \sigma^{3} &             \\
             & -\sigma^{3} \\  
  \end{array}
\right)\, ,
\hspace{.5cm}
(T_{2} {}^{M}{}_{N})
= 
-\tfrac{1}{2} 
\left( 
  \begin{array}{cc}
           &   \sigma^{3} \\
\sigma^{3} &              \\  
  \end{array}
\right)\, ,
\hspace{.5cm}
(T_{3}{}^{M}{}_{N})
= 
\tfrac{1}{2} 
\left( 
  \begin{array}{cc}
             & \mathbbm{1} \\
-\mathbbm{1} &             \\  
  \end{array}
\right)\, .
\end{equation}

The same transformations act on all the symplectic vectors of the
theory and, in particular, on the variables $H^{M}$ and the charge
vectors $\mathcal{Q}^{M}$. In this formulation of the axidilaton
system there seem to be no further
symmetries\footnote{\label{foot:marrani} We will see, however, that
  there is an additional U$(1)$ factor in the symmetry group that only
  has a non-trivial action on objects with symplectic indices and that
  coincides with the continuous global Freudenthal duality
  transformation. The scalars do not transform under this symmetry. On
  the other hand, only this U$(1)$ symmetry is also a local symmetry
  of the H-FGK formalism. We would like to thank Alessio Marrani for
  clarifying discussions on this point.}.


\subsubsection{Equivariant vectors of the axidilaton model}
\label{sec-equiaxi}


In this model there is no need to solve any equation to find 4
linearly independent equivariant vectors:  observe that the symplectic
vector of charges is the direct sum of two real $\mathrm{Sl}(2;\mathbb{R})$
doublets $a^{i}$ and $b_{i}$ ($i,j=1,2$), namely

\begin{equation}
(a^{i}) 
\equiv
\left(
\!\!
  \begin{array}{c}
p^{1} \\ q_{1} \\    
  \end{array}
\!\!
\right)
\, ,
\hspace{1.5cm}  
(b_{i})
\equiv 
(p^{0},\, q_{0})\, . 
\end{equation}

\noindent
These doublets transform respectively contravariantly and covariantly, that is

\begin{equation}
\hspace{1cm}  
a^{\prime\, i}
=
\Lambda^{i}{}_{j}\, a^{j}\, ,
\hspace{1cm}  
b^{\prime}_{i}
=
b_{j}\, 
(\Lambda^{-1})^{j}{}_{i}\, ,
\end{equation}

\noindent
where $(\Lambda^{i}{}_{j})$ is the matrix given in
Eq.~(\ref{eq:SL2Rmatrix}), which furthermore satisfies

\begin{equation}
(\Lambda^{-1})^{i}{}_{j}= \Omega^{ki}\, \Lambda^{l}{}_{k}\, \Omega_{lj}\, ,   
\hspace{1cm}
(\Omega_{ij}) = (\Omega^{ij})= 
\left(
  \begin{array}{cc}
  0 & 1 \\ -1 & 0 \\  
  \end{array}
\right)\, ,
\end{equation}

\noindent
because $\mathrm{Sl}(2;\mathbb{R})\simeq \mathrm{Sp}(2;\mathbb{R})$. We can use the symplectic
metric $\Omega$ to raise and lower doublet indices such as $i$ and $j$  using
the same convention we use for the symplectic indices (see footnote \ref{foot:1}), namely $a_{i}\equiv
\Omega_{ij}a^{j}$ and $b^{i}= b_{j}\Omega^{ji}$.  The only non-vanishing
$\mathrm{Sl}(2;\mathbb{R})$ invariant that can be built out of these two doublets is

\begin{equation}
a^{i}b_{i}
=
p^{0}p^{1} + q_{0}q_{1}
\equiv
\tfrac{1}{2}
\mathsf{W}(\mathcal{Q})\, .
\end{equation}

Let us denote by $\mathcal{Q}^{M}(a,b)$ the standard symplectic charge
vector seen as the direct sum of the two doublets $a$ and $b$. 
Using the two doublets we can construct three further, up to a global sign,
inequivalent charge vectors that under $\mathrm{Sl}(2;\mathbb{R})$ transform
in the same way as $\mathcal{Q}^{M}(a,b)$, {\em i.e.\/} equivariantly; the
four equivariant charge vectors are

\begin{equation}
\label{eq:QMab}
  \begin{array}{rclrcl}
\mathcal{Q}^{M}(a,b) 
& \equiv &
\left( \!
  \begin{array}{c}
p^{0} \\ p^{1} \\ q_{0} \\ q_{1} \\    
  \end{array}
\!
\right)\, ,
\hspace{1cm}
&
\mathcal{Q}^{M}(b,-a) 
& \equiv &
\left( \!
  \begin{array}{c}
-q_{1} \\ -q_{0} \\ p^{1} \\ p^{0} \\    
  \end{array}
\!
\right)\, ,  
\\
& & & & & \\
\mathcal{Q}^{M}(-a,b) 
& \equiv &
\left( \!
  \begin{array}{c}
p^{0} \\ -p^{1} \\ q_{0} \\ -q_{1} \\    
  \end{array}
\!
\right)\, ,  
\hspace{1cm}
&
\mathcal{Q}^{M}(-b,-a) 
& \equiv &
\left( \!
  \begin{array}{c}
-q_{1} \\ q_{0} \\ p^{1} \\ -p^{0} \\    
  \end{array}
\!
\right)\, .  
\\
\end{array}
\end{equation}

These equivariant vectors are generically linearly independent and provide a
basis of equivariant vectors; any other equivariant
vector, in particular the attractors $B^{M}$, can be expanded w.r.t.~this base, {\em e.g.\/}

\begin{equation}
B^{M} = b^{\sigma} U_{\sigma}^{M}\, ,
\,\,\,\,\,
\mathrm{with}
\,\,\,\,\,
\{U_{\sigma}\} = \{\mathcal{Q},\tilde{\mathcal{Q}},\mathcal{P},\tilde{\mathcal{P}}\}\, .
\end{equation}

We will plug this general ansatz into the equation $\left.\partial_{M}V_{\rm
    bh}(H,\mathcal{Q})\right|_{H=B}=0$ as to find the most general attractor of the
theory in Section~\ref{eq:criticalpointsaxidilaton}, but at this point we already know some
general results: The standard charge vector $\mathcal{Q}^{M}(a,b)$ will be the
supersymmetric attractor, as usual, and we are going to see,
$\mathcal{Q}^{M}(b,-a)$ is its Freudenthal dual

\begin{equation}
\mathcal{Q}^{M}(b,-a) 
= 
\tilde{\mathcal{Q}}^{M}(a,b)= \tilde{\mathcal{Q}}^{M}\, .  
\end{equation}

\noindent
On the other hand, $\mathcal{Q}^{M}(-a,b)$ is the non-supersymmetric
attractor $\mathcal{P}^{M}$ and $\mathcal{Q}^{M}(b,a)$ is its Freudenthal dual

\begin{equation}
\mathcal{Q}^{M}(-a,b)
=
\mathcal{P}^{M}\, ,
\hspace{1cm}
\mathcal{Q}^{M}(b,a) 
= 
\tilde{\mathcal{Q}}^{M}(b,-a) = \tilde{\mathcal{P}}^{M}\, .  
\end{equation}

\noindent
It is easy to see that

\begin{equation}
\mathsf{W}(\tilde{\mathcal{Q}}) 
= 
\mathsf{W}(\mathcal{Q}) 
= 
-\mathsf{W}(\mathcal{P}) 
= 
-\mathsf{W}(\tilde{\mathcal{P}})\, . 
\end{equation}

These four vectors are related by $\mathrm{Sp}(4;\mathbb{R})$ transformations
that however do not belong to $\mathrm{Sl}(2;\mathbb{R})\subset \mathrm{Sp}(4;\mathbb{R})$:

\begin{eqnarray}
\tilde{\mathcal{Q}}^{M} 
& = & 
\mathcal{A}^{M}{}_{N} \mathcal{Q}^{N}\, , 
\hspace{1.5cm}
(\mathcal{A}^{M}{}_{N})
\equiv
\left(
  \begin{array}{cc}
0 & \sigma^{1} \\
-\sigma^{1} & 0 \\
  \end{array}
\right)\, ,
\\  
& & \nonumber \\
\label{eq:PMdef}
\mathcal{P}^{M} 
& = & 
 \mathcal{B}^{M}{}_{N} \mathcal{Q}^{N}\, , 
\hspace{1.5cm}
(\mathcal{B}^{M}{}_{N})
\equiv 
\left(
  \begin{array}{cc}
 \sigma^{3} & 0 \\
0 & \sigma^{3} \\   
  \end{array}
\right)\, .
\end{eqnarray}

The only non-vanishing symplectic contractions between these four vectors are 

\begin{equation}
\mathcal{\tilde{Q}}_{M}\mathcal{Q}^{M}
=
-\mathcal{\tilde{P}}_{M}\mathcal{P}^{M} 
=
\mathsf{W}(\mathcal{Q})\, .
\end{equation}

Apart from these moduli-independent equivariant vectors we can
construct the generic moduli-dependent ones by taking the real and
imaginary parts of
Eq.~(\ref{eq:genericmodulidependentequivariantvector}), in which we
can replace $\mathcal{Q}$ by any of the other three equivariant
vectors. Observe that when we use the Freudenthal dual charge, we obtain the same
complex equivariant vector but multiplied by $-i$.


\subsection{H-FGK formalism}


The solution of the  stabilization equations of this theory is

\begin{equation}
\mathcal{R}_{M}(\mathcal{I}) = \mathcal{A}_{MN}\mathcal{I}^{N}\, ,
\hspace{1cm}
(\mathcal{A}_{MN})
\equiv 
\left(
  \begin{array}{cc}
\sigma^{1} & 0\\
0 & \sigma^{1} \\
  \end{array}
\right)\, ,
\end{equation}

\noindent
where $\sigma^{1}$ is the standard Pauli matrix.  $\mathcal{A} =
(\mathcal{A}_{MN})$ is a symplectic matrix:

\begin{equation}
\mathcal{A}\Omega  \mathcal{A}  = \Omega\, ,
\end{equation}

\noindent
which is not surprising since it is just $-\mathcal{M}_{MN}(\mathcal{F})$.  It
follows that $(\mathcal{A}^{M}{}_{N}) = (\Omega^{PM}\mathcal{A}_{PN}) =
-\Omega\mathcal{A}$ is also a symplectic matrix.

By definition, the original and tilded, {\em i.e.\/} Freudenthal dual,
$H$-variables are related by\footnote{Explicitly, we have
  \begin{equation}
  (\tilde{H}^{M}) =
\left(
  \begin{array}{c}
-\sigma^{1\, \Lambda\Sigma}H_{\Sigma} \\
\sigma^{1}{}_{\Lambda\Sigma}H^{\Sigma} \\
\end{array}
\right)
=
\left(
  \begin{array}{c}
   -H_{1} \\  -H_{0} \\  H^{1} \\ H^{0} \\ 
  \end{array}
\right)\, .  
  \end{equation}
This vector should be compared with $\mathcal{Q}^{M}(b,-a)$ in Eq.~(\ref{eq:QMab}).
}

\begin{equation}
\tilde{H}_{M}(H) = \mathcal{A}_{MN} H^{N}\, ,
\hspace{1cm}
\tilde{H}^{M}(H) = \mathcal{A}^{M}{}_{N} H^{N}\, .
\end{equation}

\noindent
Therefore in this simple model the Freudenthal duality transformation is
linear and is, furthermore, a symplectic transformation. It is clearly a
transformation that does not belong to the global symmetries that act on the axidilaton (i.e.~$\mathrm{Sl}(2;\mathbb{R})$ whose embedding into $\mathrm{Sp}(4;\mathbb{R})$ is given in
Eq.~(\ref{eq:sl2embeddedinsp4})), but it is a symmetry transformation that
acts on objects with symplectic indices such as the vector fields and as such
must be considered a part of the duality group of the
model\footnote{See footnote~\ref{foot:marrani}.}.

As expected in Freudenthal duality

\begin{equation}
\mathcal{A}^{M}{}_{P} \ \mathcal{A}^{P}{}_{N}  = - \delta^{M}{}_{N}\, .
\end{equation}

We can extend the Freudenthal duality transformation to all symplectic
vectors. The properties

\begin{equation}
\label{eq:properties}
\tilde{X}_{M}Y^{M}
= 
\tilde{Y}_{M}X^{M}
=
-Y_{M}\tilde{X}^{M}
\, ,
\,\,\,\,\,
\Rightarrow
\,\,\,\,\,
\tilde{X}_{M}\tilde{Y}^{M}
=
X_{M}Y^{M}\, ,
\end{equation}

\noindent
which hold in this particular model for any two symplectic vectors $X^{M}$ and $Y^{M}$
because Freudenthal duality is a symplectic transformation, will be used very often.

The Hesse potential is given by the $\mathrm{Sl}(2;\mathbb{R})$ invariant
discussed in earlier sections

\begin{equation}
\label{eq:Hesseaxidilaton}
\mathsf{W}(H) 
\equiv 
\tilde{H}_{M}(H) H^{M} 
= 
\mathcal{A}_{MN}H^{M}H^{N}
=
2(H^{0}H^{1}+H_{0}H_{1})\, ,  
\end{equation}

\noindent
and in accordance with the general formalism it determines the
model completely:  the effective action can be constructed entirely from it and
the metric function $e^{-2U}$ and the axidilaton $\lambda$ are
related to the Hesse potential by

\begin{equation}
\label{eq:axidilaton}
e^{-2U} 
=
\mathsf{W}(H) \, ,
\hspace{1cm}
\lambda \equiv iZ 
=
i\frac{\tilde{H}^{1}+iH^{1}}{\tilde{H}^{0}+iH^{0}}
=
\frac{H^{1}+iH_{0}}{H_{1}-iH^{0}}\, .
\end{equation}






The metric $g_{MN}(H)$ of this system can be written in the form

\begin{equation}
\label{eq:metricaxidilaton}
g_{MN} = 2\ \mathfrak{N}_{MNPQ}\ \frac{H^{P}H^{Q}}{\mathsf{W}^{2}}\, ,  
\end{equation}

\noindent
where we have defined the constant matrix

\begin{equation}
\mathfrak{N}_{MNPQ}
\equiv
\mathcal{A}_{MN}\mathcal{A}_{PQ}  
-2\mathcal{A}_{MP}\mathcal{A}_{NQ}  
-\Omega_{MP}\Omega_{NQ}\, .  
\end{equation}

\noindent
Using this notation, the derivatives of the metric take the form

\begin{equation}
\partial_{M}g_{PQ}
=  
-4 \frac{\tilde{H}_{M}}{\mathsf{W}}g_{PQ}
+4 \mathfrak{N}_{PQ(MR)} \frac{H^{R}}{\mathsf{W}^{2}}\, , 
\end{equation}

\noindent
and the Christoffel symbols of the first kind are given by\footnote{We
  remind the reader that the metric $g_{MN}(H)$ is not invertible, so
  we cannot use the standard Christoffel symbols $ \Gamma_{PQ}{}^{M}\equiv g^{NM}[PQ,M]$.}

\begin{equation}
  \begin{array}{rcl}
[PQ,M]
& = &
2 {\displaystyle\frac{\tilde{H}_{M}g_{PQ}
  -\tilde{H}_{P}g_{QM}-\tilde{H}_{Q}g_{PM}}{\mathsf{W}}}
\\
& & \\
& & 
-[6\mathcal{A}_{PQ}  \mathcal{A}_{MR}
-4\mathcal{A}_{M(P}\mathcal{A}_{Q)R}  
+4\Omega_{M(P}\Omega_{Q)R}]
{\displaystyle\frac{H^{R}}{\mathsf{W}^{2}}}\, . 
\end{array}
\end{equation}

\noindent
It is easy to check that $\tilde{H}^{M}[PQ,M]=0$, as required by
Freudenthal duality invariance.

The potential $V$ can be written in the convenient
form 

\begin{equation}
\label{eq:potentialaxidil}
\mathsf{W}^{2}V(H,\mathcal{Q})
=
-\tfrac{1}{2}\mathsf{W}(\mathcal{Q})\mathsf{W}
+(H^{M}\tilde{\mathcal{Q}}_{M})^{2}
+(H^{M}\mathcal{Q}_{M})^{2}
\, ,
\end{equation}

\noindent
and its derivative reads

\begin{equation}
\label{eq:derivativepotentialaxidil}
\partial_{M}V
=
-4\frac{\tilde{H}_{M}}{\mathsf{W}}
\left[V 
+\tfrac{1}{4}\frac{\mathsf{W}(\mathcal{Q})}{\mathsf{W}}
\right]
+2(\mathcal{Q}_{M}\mathcal{Q}_{N}
+\tilde{\mathcal{Q}}_{M}\tilde{\mathcal{Q}}_{N})
\frac{H^{N}}{\mathsf{W}^{2}}\, ;
\end{equation}

\noindent
using the
properties Eq.~(\ref{eq:properties}) it is easy to see that $\tilde{H}^{M}\partial_{M}V =0$, which is the last requirement
for having local Freudenthal duality \cite{Galli:2012ji}.

Observe that, in this model, a Freudenthal duality transformation of the
charge vectors \textit{only} (that is: not of the variables $H^{M}$), not only
preserves $\mathsf{W}(\mathcal{Q})$ but also the complete potential and
black-hole potential, {\em i.e.\/}

\begin{equation}
\mathsf{W}(\tilde{\cal Q}) = \mathsf{W}(\mathcal{Q})
\,\,\,\, 
\Rightarrow
\,\,\,\,
V(H,\tilde{\mathcal{Q}})
=
V(H,\mathcal{Q})\, ,
\,\,\,\,\,
\text{and}
\,\,\,\,\,
V_{\rm bh}(H,\tilde{\mathcal{Q}})
=
V_{\rm bh}(H,\mathcal{Q})
\, .
\end{equation}



On the other hand, using the definition of the fake charge
Eq.~(\ref{eq:fakecentralcharge1}) one can show that for any values of
$H^{M}$

\begin{eqnarray}
\label{eq:V=Z+Ztris}
-V_{\rm bh}(\mathcal{Q})
& = &
-\tfrac{1}{2}\mathsf{W}(\mathcal{Q})
+2|\mathcal{Z}(\mathcal{Q})|^{2}
=
-\tfrac{1}{2}\mathsf{W}(\mathcal{P})
+2|\mathcal{Z}(\mathcal{P})|^{2}
=
-V_{\rm bh}(\mathcal{P})
\, , 
\\ 
& & \nonumber \\
|\mathcal{Z}(\mathcal{P})|^{2}
& = & 
|\mathcal{Z}(\mathcal{Q})|^{2}
-\tfrac{1}{2}\mathsf{W}(\mathcal{Q})\, .
\\
\end{eqnarray}

\noindent
The first identity means that, if $\mathcal{Q}$ is an attractor, so
will $\mathcal{P}$. The fact that it is an identity for arbitrary
values of $H^{M}$ means that replacing $\mathcal{Q}$ by $\mathcal{P}$
in an extremal solution gives another extremal solution with the
attractor $\mathcal{P}$. The second identity is a consequence of the
first and implies that

\begin{equation}
  \begin{array}{rclcrcl}
\mathsf{W}(\mathcal{Q}) & < & 0\, , & \Rightarrow &  
|\mathcal{Z}(\mathcal{P})| &  > & |\mathcal{Z}(\mathcal{Q})|\, ,\\
& & & & & & \\
\mathsf{W}(\mathcal{Q}) & > & 0\, , & \Rightarrow &  
|\mathcal{Z}(\mathcal{Q})| &  > & |\mathcal{Z}(\mathcal{P})|\, ,\\
  \end{array}
\end{equation}

\noindent
for all values of $H^{M}$. The second case should correspond to the
supersymmetric attractor in which the evaporation process stops when
the mass equals the largest central charge, which in this case is the
true one.

Finally, observe that this black-hole potential satisfies the curious
interchange property

\begin{equation}
V_{\rm bh}(H,\mathcal{Q})
= \frac{\mathsf{W}(H)}{\mathsf{W}(\mathcal{Q})} \
V_{\rm bh}(\mathcal{Q},H)\, .
\end{equation}


\subsection{The symmetries in the H-FGK formalism}


In Section~\ref{sec-globalsymaxidil} we discussed the global symmetries of the
axidilaton model (more precisely, of its scalar manifold metric) when it is
described in terms of the standard fields and have studied the embedding of
these symmetries into $\mathrm{Sp}(4;\mathbb{R})$. It is in this form that we expect
these symmetries to be present in the H-FGK formalism. On the other hand,
there may be additional non-obvious symmetries such as Freudenthal duality (which is in
general non-linear) in the H-FGK formalism.

Let us consider first the kinetic term: if we consider only linear
transformations of the $H^{M}$

\begin{equation}
\delta H^{M} = T^{M}{}_{N} H^{N}\, ,  
\end{equation}

\noindent
it is evident that they will leave the kinetic term invariant if they are symplectic
transformations, {\em i.e.\/}

\begin{equation}
\Omega_{P(M}T^{P}{}_{N)}= 0\, ,
\end{equation}

\noindent
and are furthermore symmetries of the Hesse potential

\begin{equation}
\delta \mathsf{W} = 2 \tilde{H}_{M} \delta H^{M} = 2\tilde{H}_{M}\ T^{M}{}_{N}\
H^{N} =0
\,\,\,\,\,\,
\longrightarrow
\,\,\,\,\,\,
[\Omega\mathcal{A},T] = \beta \mathbbm{1}_{4\times 4}\, ,  
\end{equation}

\noindent
where $\beta$ is a real constant that can vanish. It is not difficult to see
that for infinitesimal symplectic transformations, $\beta$ must indeed
vanish, and the only independent generators that solve the above equation are
the three $\mathrm{Sl}(2;\mathbb{R})$ generators $T_{i}$ given in
Eq.~(\ref{eq:symplecticTi}) plus

\begin{equation}
T_{4}= \tfrac{1}{2}\mathcal{A}\Omega\, ,  
\end{equation}

\noindent
which generates the Freudenthal transformations and commutes with the generators
of $\mathrm{Sl}(2;\mathbb{R})$\footnote{it is not difficult to see that the Hesse
  potential of the axidilaton model is not determined by $\mathrm{Sl}(2;\mathbb{R})$
  invariance alone: one must require invariance under Freudenthal
  duality.}.

It can be checked that these symmetries leave invariant the metric
$g_{MN}$. Actually, the metric is invariant under the constant rescalings of
the $H^{M}$

\begin{equation}
T_{5} \equiv \tfrac{1}{4} \mathbbm{1}_{4\times 4}\, ,
\end{equation}

\noindent
which are not symplectic transformations and leave the Hesse potential
invariant only up to a multiplicative constant, in the same way as the K\"ahler potential
is invariant under isometries of the K\"ahler metric only up to K\"ahler
transformations.

We can study now the invariance of the potential using the expression for
$\partial_{M}V$ given in Eq.~(\ref{eq:derivativepotentialaxidil}). The
first term cancels for $i=1,2,3,4$ (we do not need to check $i=5$: the
potential is homogeneous of degree $-2$ and $\delta_{5}V=-2V \neq 0$ in
general) and the rest gives

\begin{equation}
\delta_{i}V
=
-2H^{N}T_{i}{}^{M}{}_{N}(\mathcal{Q}_{M}\mathcal{Q}_{N}
+\tilde{\mathcal{Q}}_{M}\tilde{\mathcal{Q}}_{N})
\frac{H^{N}}{\mathsf{W}^{2}}\, ,
\end{equation}

\noindent
which vanishes only for the Freudenthal transformation $i=4$ unless
we also perform the same transformation on the charge vector: this
means that $\mathrm{Sl}(2;\mathbb{R})$ is only a \textit{pseudo-symmetry} of
the system, since the constants that enter the action are rotated. The
charges appear as integration constants of the solution of the
equations of motion for the electrostatic and magnetostatic potentials
in Ref.~\cite{Ferrara:1997tw} and $\mathrm{Sl}(2;\mathbb{R})$ is probably a
(standard) symmetry of the effective theory before that. 

There are no conserved quantities associated to pseudo-symmetries, whence
there is only one conserved current: the one associated to the
Freudenthal duality. This current vanishes, however, identically, which is a
generic feature of the formalism.

\subsection{Critical points}
\label{eq:criticalpointsaxidilaton}



The critical points of this model are equivariant vectors $B^{M}$ satisfying
the equations

\begin{equation}
\left.\partial_{M}V_{\rm bh} \right|_{H=B}
=
-2\frac{\tilde{B}_{M}}{\mathsf{W}(B)}
\left[V_{\rm bh}(B,\mathcal{Q}) 
-\tfrac{1}{2}\mathsf{W}(\mathcal{Q})
\right]
-2(\mathcal{Q}_{M}\mathcal{Q}_{N}
+\tilde{\mathcal{Q}}_{M}\tilde{\mathcal{Q}}_{N})
\frac{B^{N}}{\mathsf{W}(B)}=0\, .
\end{equation}

Using the basis of equivariant vectors $\{U_{\sigma}\} =
\{\mathcal{Q},\tilde{\mathcal{Q}},\mathcal{P},\tilde{\mathcal{P}}\}$
constructed in Section~\ref{sec-equiaxi}, we can write any such solution as

\begin{equation}
B^{M} 
= 
a \mathcal{Q}^{M} +\tilde{a} \tilde{\mathcal{Q}}^{M}   
+
b \mathcal{P}^{M} +\tilde{b} \tilde{\mathcal{P}}^{M}\, .
\end{equation}

The only non-vanishing symplectic products of the four basis vectors are 

\begin{equation}
\tilde{\mathcal{Q}}_{M} \mathcal{Q}^{M} = \mathsf{W}(\mathcal{Q})\, ,
\hspace{1cm}
\tilde{\mathcal{P}}_{M} \mathcal{P}^{M} = -\mathsf{W}(\mathcal{Q})\, ,
\end{equation}

\noindent
and a very simple calculation gives

\begin{equation}
  \begin{array}{rcl}
\left.\partial_{M} V_{\rm bh}\right|_{H=B}
& = & 
{\displaystyle\frac{-2}{(a^{2}+\tilde{a}^{2} -b^{2} -\tilde{b}^{2})}}
\left\{
\tilde{a}( b^{2} +\tilde{b}^{2}) \mathcal{Q}_{M} 
-a( b^{2} +\tilde{b}^{2})\tilde{\mathcal{Q}}_{M}   
\right.
\\
& & \\
& & 
\left.
+
\tilde{b}( a^{2} +\tilde{a}^{2}) \mathcal{P}_{M} 
-b ( a^{2} +\tilde{a}^{2})\tilde{\mathcal{P}}_{M}
\right\}=0\, ,\\
\end{array}
\end{equation}

\noindent
which only admits two non-trivial solutions: $b=\tilde{b}=0$ and
$a=\tilde{a}=0$.  The first solution, up to global normalization (which is
undetermined in this formalism because the black-hole potential is
scale-invariant), corresponds to a global Freudenthal rotation with arbitrary
angle of the standard supersymmetric attractor $B^{M}=\mathcal{Q}^{M}$ and the
second corresponds to a global Freudenthal rotation of the standard
non-supersymmetric attractor $B^{M}=\mathcal{P}^{M}$ \cite{Galli:2011fq}.

We obtain the following relations

\begin{equation}
\label{eq:normalizations}
V_{\rm bh}(\mathcal{P},\mathcal{P}) 
=
-V_{\rm bh}(\mathcal{Q},\mathcal{P})
=
V_{\rm bh}(\mathcal{P},\mathcal{Q})
=
-V_{\rm bh}(\mathcal{Q},\mathcal{Q})
=
\tfrac{1}{2}\mathsf{W}(\mathcal{Q})
\, ,
\end{equation}

\noindent
that are necessary to have the corresponding near-horizon solutions, see
Eq.~(\ref{eq:conditionNHsolution}).


\subsection{Conventional extremal solutions}


As a first simple illustration of the methods proposed in the first section of
this paper, we are going to review the construction of the extremal
solutions\footnote{The axidilaton model is a particular case ($n=1$) of the
  $\overline{\mathbb{CP}}^{n}$ model. We will construct the most general
  non-extremal solutions of that model (and, hence, of the axidilaton model)
  later.} performed in Ref.~\cite{Galli:2012pt}. 

From the results of that paper we know that all of them (including the
extremal non-supersymmetric ones) are going to be conventional, but it is
important for us to understand why. Thus, we start from the near-horizon
solutions given by Eq.~(\ref{eq:NHsolution}) where $B^{M}$ takes the values of
the attractors found in the previous section, normalized so that (see
Eq.~(\ref{eq:conditionNHsolution}))

\begin{equation}
V_{\rm bh}(B,\mathcal{Q}) =  V_{\rm bh}(B,B) = -\tfrac{1}{2}\mathsf{W}(B)\, .
\end{equation}

\noindent
The attractors that satisfy these conditions are global Freudenthal rotations
of the standard supersymmetric attractor $\mathcal{Q}^{M}$ and of the
non-supersymmetric one $\mathcal{P}^{M}$, {\em i.e.\/} 

\begin{eqnarray}
  \label{eq:1}
  \mbox{either}\;\;  &B^{M} \ =& \cos{\theta}\, \mathcal{Q}^{M} + \sin{\theta}\, \tilde{\mathcal{Q}}^{M} 
      \; ,\nonumber \\ & & \nonumber \\
  \mbox{or~~~~~~}\;\;  &B^{M} \ =& \cos{\theta}\, \mathcal{P}^{M} + \sin{\theta}\, \tilde{\mathcal{P}}^{M}\, .
\end{eqnarray}

\noindent
The results of Section~(\ref{sec-conventional}) guarantee that
Eq.~(\ref{eq:NHsolution}) provides a near-horizon solution for these choices
of $B^{M}$. Now, to see if we can extend these solutions to asymptotically
flat solutions by adding an infinitesimal constant vector to these $H^{M}$ as
in Eq.~(\ref{eq:perturbedNHsolution}), we have to compute the rank of
$\mathfrak{M}_{MN}$ in Eq.~(\ref{eq:conditioninfinitesimalextension}) to find
how many independent solutions $\varepsilon^{M}$ exist.

It is enough to consider a charge configuration whose orbit covers the
complete charge space (see Appendix~\ref{app:generating}) and, therefore, we
set $p^{0}=p^{1}=0$, getting, for the supersymmetric ($+$) and
non-supersymmetric ($-$) cases, the matrix

\begin{equation}
  (\mathfrak{M}_{MN})
=
\tfrac{1}{2} 
\left(
  \begin{array}{cccc}
  \frac{1}{q_{1}^{2}} & \pm\frac{1}{q_{0}q_{1}} & 0 & 0 \\
 \pm \frac{1}{q_{0}q_{1}} &  \frac{1}{q_{0}^{2}} & 0 & 0 \\
  0 & 0 & 0 & 0 \\
  0 & 0 & 0 & 0 \\
  \end{array}
\right)\, .
\end{equation}

\noindent
This matrix has rank 1 and, furthermore, the three independent solutions to
Eq.~(\ref{eq:conditioninfinitesimalextension}) satisfy the constraint
$B^{M}\varepsilon_{M}=0$. This means that there is no obstruction to the
addition of arbitrary (up to normalization $\mathsf{W}(A)=1$ and the condition
$B^{M}A_{M}=0$) constants $A^{M}$ to the near-horizon harmonic functions,
which now take the form

\begin{equation}
H^{M} = A^{M} -\tfrac{1}{\sqrt{2}} B^{M} \tau\, .  
\end{equation}

The two independent components of $A^{M}$ describe the two real moduli of this
theory $\Re\mathfrak{e}(\lambda_{\infty})\,,$ $\Im\mathfrak{m}(\lambda_{\infty})$
and $A^{M}$ is given by \cite{Galli:2012pt}

\begin{equation}
\label{eq:AMextremal}
A^{M} = \sqrt{2} \, 
\Im\mathfrak{m} 
\left(\frac{\mathcal{Z}^{*}(\phi_{\infty},B)}{|\mathcal{Z}(\phi_{\infty},B)|}
\mathcal{V}^{M}_{\infty}\right)\, .  
\end{equation}

To show that the equations of motion are satisfied for finite constants
$A^{M}$ (which is only needed in the non-supersymmetric case) we can proceed
as follows: from the linearity of the $H^{M}$ it is possible to show that these
configurations satisfy first-order flow equations \cite{Ortin:2011vm}. These,
in turn can be shown to imply the standard second-order equations of motion if
and only if the identity

\begin{equation}
V_{\rm bh}(H,\mathcal{Q}) =   V_{\rm bh}(H,B)\, ,   
\end{equation}

\noindent
is satisfied for arbitrary values of $H$. This is evident for
$B^{M}=\mathcal{Q}^{M}$ (the supersymmetric attractor) and has been shown for
$B^{M}=\mathcal{P}^{M}$ (the non-supersymmetric attractor) in
Eq.~(\ref{eq:V=Z+Z}) and the invariance of the black-hole potential under
Freudenthal transformations of the charges extends this result to the other
two (physically indistinguishable) attractors and proves that these
configurations are classical solutions of the model.


\subsection{Unconventional solutions}


We do not expect more extremal black-hole solutions to the axidilaton model
since the solutions constructed in the previous section already have the
maximal number of independent physical parameters (charges $\mathcal{Q}^{M}$
and moduli $\lambda_{\infty}$) which are constrained only by the requirement
that the horizon has a non-vanishing area, {\em i.e.\/} $\mathsf{W}(B)>0$.

On the other hand, we can rewrite these solutions in an unconventional form
(\textit{i.e.}~so that $\dot{H}^{M}H_{M}\neq 0$) by using local Freudenthal
duality transformations, but in this case doing so merely complicates the
form of the solution in the H-FGK formalism.


\section{The $\overline{\mathbb{CP}}^{n}$ model}
\label{sec-CPn}


The prepotential of the $\overline{\mathbb{CP}}^{n}$ model is given
by\footnote{The black-hole solutions of this model have been studied in
  \cite{Galli:2011fq}.}

\begin{equation}
\mathcal{F} 
=
-\tfrac{i}{4}\eta_{\Lambda\Sigma}
\mathcal{X}^{\Lambda}\mathcal{X}^{\Sigma}\, ,
\hspace{1cm}
(\eta_{\Lambda\Sigma}) = \mathrm{diag}(+-\dotsm -)\, .
\end{equation}

\noindent
The $\overline{\mathbb{CP}}^{n}$ model contains $n$ scalar fields given by

\begin{equation}
Z^{i} \equiv \mathcal{X}^{i}/\mathcal{X}^{0}\, ,
\end{equation}

\noindent
but it is convenient to add $Z^{0}\equiv 1$ and we define

\begin{equation}
(Z^{\Lambda})
\equiv
\left(\mathcal{X}^{\Lambda}/\mathcal{X}^{0}\right) = (1,Z^{i})\, ,
\hspace{1cm}
(Z_{\Lambda})
\equiv
(\eta_{\Lambda\Sigma}Z^{\Sigma})= (1,Z_{i}) =(1,-Z^{i})\, .
\end{equation}

\noindent
The K\"ahler potential, the K\"ahler metric, the inverse K\"ahler metric and
the covariantly holomorphic symplectic section read

\begin{equation}
\label{Vcpn}
  \begin{array}{rcl}
\mathcal{K} 
& = & 
-\log{(Z^{*\Lambda}Z_{\Lambda})}\, ,
\\
& & \\
\mathcal{G}_{ij^{*}} 
& = & 
-e^{\mathcal{K}}\left( \eta_{ij^{*}}
-e^{\mathcal{K}}Z^{*}_{i}Z_{j^{*}} \right)\, ,
\\
& & \\
\mathcal{G}^{ij^{*}} 
& = &
-e^{-\mathcal{K}} \left(\eta^{ij^{*}}
+Z^{i}Z^{*\, j^{*}}\right)\, ,
\\
& & \\
\mathcal{V}
& = &
e^{\mathcal{K}/2}
\left(
  \begin{array}{c}
  Z^{\Lambda} \\ \\ -\tfrac{i}{2} Z_{\Lambda} \\
  \end{array}
\right)\, .
\end{array}
\end{equation}

It is also convenient to define the following complex charge combinations

\begin{equation}
\label{charg}
\Gamma_{\Lambda}
\equiv
q_{\Lambda} +\tfrac{i}{2}\eta_{\Lambda\Sigma}p^{\Sigma}\, ,
\end{equation}

\noindent
in terms of which the central charge, its holomorphic K\"ahler-covariant
derivative and the black-hole potential are

\begin{equation}
\label{eq:centralchargescpn}
\begin{array}{rcl}
\mathcal{Z}
& = &
e^{\mathcal{K}/2}Z^{\Lambda}\Gamma_{\Lambda} 
\equiv \mathcal{Z}(\Gamma)\, ,
\\
& & \\
\mathcal{D}_{i}\mathcal{Z}
& = &
e^{3\mathcal{K}/2}Z^{*}_{i}Z^{\Lambda}\Gamma_{\Lambda}
-e^{\mathcal{K}/2}\Gamma_{i}\, ,
\\
& & \\
-V_{\rm bh}
& = &
2e^{\mathcal{K}}|Z^{\Lambda}\Gamma_{\Lambda}|^{2}
-\Gamma^{*\, \Lambda}\Gamma_{\Lambda}\, .
\end{array}
\end{equation}

We can extend this complex notation to any symplectic vector:

\begin{equation}
\mbox{if}\;\;\;
(A^{M}) 
= 
\left(
\begin{array}{c}
    a^{\Lambda} \\ b_{\Lambda} \\
\end{array}
\right)
\,\,\,\,\,
\mbox{then}
\hspace{.6cm}
\left\{\begin{array}{lcl}
\mathcal{A}_{\Lambda} & \equiv& 
b_{\Lambda}+\tfrac{i}{2}\eta_{\Lambda\Sigma}a^{\Sigma} \; ,\\
  & & \\
\mathcal{A}^{\Lambda} &\equiv &\eta^{\Lambda\Sigma} \mathcal{A}_{\Sigma}
\ =\
\eta^{\Lambda\Sigma}b_{\Sigma}+\tfrac{i}{2}a^{\Lambda}\, ,
\end{array}\right.
\end{equation}

\noindent
and the symplectic product of two vectors becomes

\begin{equation}
A_{M}B^{M} = -2\Im\mathfrak{m} (\mathcal{A}_{\Lambda}\mathcal{B}^{*\, \Lambda})\, , 
\end{equation}

\noindent
where of course $\mathcal{A}_{\Lambda}\mathcal{B}^{*\, \Lambda}=
\mathcal{A}^{\Lambda}\mathcal{B}^{*}{}_{\Lambda}$. We will use both notations, based on
convenience.


\subsection{The global symmetries of the $\overline{\mathbb{CP}}^{n}$ model}


The $n$ complex scalars of the $\overline{\mathbb{CP}}^{n}$ model parametrize
the symmetric coset space $\mathrm{SU}(1,n)/\mathrm{SU}(n)$, and the full theory is invariant
under global SU($1,n$) transformations\footnote{Actually, the coset space can
  also be described as $\mathrm{U}(1,n)/\mathrm{U}(n)$, which would imply that the global
  symmetry group of the model is $\mathrm{U}(1,n)$. As in the axidilaton model (the
  $n=1$ case), the extra $\mathrm{U}(1)$, that does not act on the scalars, is the
  Freudenthal duality group (see footnote~\ref{foot:marrani}). We thank
  Alessio Marrani for clarifying discussions on this point.}. If
$\Lambda^{\Lambda}{}_{\Sigma}$ is a generic element in the fundamental
representation of $\mathrm{SU}(1,n)$, \textit{i.e.}  if it satisfies

\begin{equation}
\Lambda^{*\, \Gamma}{}_{\Lambda}\, \eta_{\, \Gamma\Delta}\, \Lambda^{\Delta}{}_{\Sigma}
=
\eta_{\Lambda\Sigma}\, ,
\hspace{.5cm}
(\mathrm{or}\,\,\,\, \Lambda^{\dagger}\eta \Lambda=\eta) \, ,
\hspace{1cm}
\det \Lambda=1 \, ,
\end{equation}

\noindent
then its action on the scalars is given by

\begin{equation}
\label{refsca}
Z^{ \prime\, \Lambda}
=
\frac{\Lambda^{\Lambda}{}_{\Sigma}Z^{\Sigma}}{\Lambda^{0}{}_{\Sigma}Z^{\Sigma}}\,
,
\hspace{1cm}
Z^{ \prime}{}_{\Lambda}
=
\frac{\Lambda_{\Lambda}{}^{\Sigma}Z_{\Sigma}}{\Lambda^{0}{}_{\Sigma}Z^{\Sigma}} \, ,
\end{equation}

\noindent
where we have raised and lowered the indices of the $\mathrm{SU}(1,n)$ matrix with the
metric $\eta$.  In the
fundamental representation the $n(n+2)$ infinitesimal generators of $\mathfrak{su}(1,n)$

\begin{equation}
\Lambda^{\Lambda}{}_{\Sigma}
\sim
\delta^{\Lambda}{}_{\Sigma}+\alpha^{m}\ T_{m}{}^{\Lambda}{}_{\Sigma}\, ,
\end{equation}

\noindent
are matrices such that $T_{m\, \Lambda\Sigma}=
\eta_{\Lambda\Gamma}\, T_{m}{}^{\Gamma}{}_{\Sigma}$ is
anti-Hermitean. Substituting the infinitesimal linear transformations in the
non-linear transformation rules of the scalars, Eq.~(\ref{refsca}), we find that
they take the form 

\begin{equation}
Z^{\prime\, \Lambda}
=
Z^{\Lambda}+\alpha^{m} k_{m}{}^{\Lambda}(Z)\, ,
\end{equation}

\noindent
where $k_{m}{}^{\Lambda}(Z)$, the holomorphic part of the Killing vectors
$K_{m}$, is given by\footnote{The $\Lambda=0$ component vanishes, as it should,
  but it is useful to keep it.}

\begin{equation}
k_{m}{}^{\Lambda}(Z)
=
T_{m}{}^{\Lambda}{}_{\Sigma}\ Z^{\Sigma}
- T_{m}{}^{0}{}_{\Omega}\  Z^{\Omega} Z^{\Lambda}\, .   
\end{equation}

\noindent
The commutation relations of the generators $T_{m}$ and the Lie brackets of
the Killing vectors are related as usual:

\begin{equation}
[T_{m},T_{n}]= f_{mn}{}^{p}\, T_{p}\, ,
\hspace{1cm}
[K_{m},K_{n}]=-f_{mn}{}^{p}\, K_{p}\, .
\end{equation}

The action of the finite $\mathrm{SU}(1,n)$ transformations on the K\"ahler potential
and on the canonical covariantly-holomorphic symplectic section $\mathcal{V}$
are given by the obvious generalization of Eqs.~(\ref{eq:Kduality}) and
(\ref{eq:Vduality}) where now

\begin{eqnarray}
f(Z)
& = & 
\log{\left(\Lambda^{0}{}_{\Sigma} Z^{\Sigma} \right)}\, ,\\
& & \nonumber \\
(S^{M}{}_{N})
& = &
\left( 
\begin{array}{cc}
\Re\mathfrak{e}\Lambda^{\Lambda}{}_{\Sigma} 
& 
-2\Im\mathfrak{m} \Lambda^{\Lambda \Sigma} \\
& \\
\tfrac{1}{2} \Im\mathfrak{m} \Lambda_{\Lambda \Sigma} 
& 
\Re\mathfrak{e}\Lambda_{\Lambda}{}^{\Sigma} 
\end{array}
\right)\, ,
\end{eqnarray}

\noindent
where once again we have raised and lowered the indices of $\Lambda^{\Lambda}{}_{\Sigma}$
with $\eta$.  The condition $\Lambda^{\dagger}\eta \Lambda=\eta$ implies
for the real and imaginary parts of $\Lambda$

\begin{equation}
\Re\mathfrak{e} \Lambda_{\Delta \Lambda}\  \Im\mathfrak{m} \Lambda^{\Delta}{}_{\Sigma}
=
\Im\mathfrak{m} \Lambda_{\Delta \Lambda}\ \Re\mathfrak{e}
\Lambda^{\Delta}{}_{\Sigma}\, ,
\hspace{1cm}
\Re\mathfrak{e} \Lambda_{\Delta \Lambda}\  \Re\mathfrak{e} \Lambda^{\Delta}{}_{\Sigma}
+
\Im\mathfrak{m} \Lambda_{\Delta \Lambda}\  \Im\mathfrak{m} \Lambda^{\Delta}{}_{\Sigma}
= 
\eta_{\Lambda\Sigma}\, ,
\end{equation}

\noindent
and implies that the matrix $(S^{M}{}_{N})$ constructed above satisfies $S^{T}
\Omega S =\Omega$ and therefore belongs to $\mathrm{Sp}(2n+2;\mathbb{R})$.  The
infinitesimal generators in this representation, {\em i.e.\/} $(T_{m}{}^{M}{}_{N})$, can be
constructed in the same way, leading to

\begin{equation}
(T_{m}{}^{M}{}_{N})
=
\left( 
\begin{array}{cc}
\Re\mathfrak{e}T_{m}{}^{\Lambda}{}_{\Sigma} 
& 
-2 \Im\mathfrak{m} T_{m}{}^{\Lambda \Sigma}
\\
& \\
\tfrac{1}{2}\Im\mathfrak{m}T_{m\, \Lambda \Sigma} 
& 
\Re\mathfrak{e}T_{m\, \Lambda}{}^{\Sigma}
\\
\end{array}
\right)\, .
\end{equation}


\subsubsection{Equivariant vectors}
\label{sec111}


The search for equivariant vectors is simplified by using the complex
combinations defined above: we look for vectors $\mathcal{B}^{\Lambda}$
behaving as $\Gamma^{\Lambda}$ under duality transformations,
\textit{i.e.}~such that its complex conjugate transforms in the fundamental
representation of $\mathrm{SU}(1,n)$

\begin{equation}
\Gamma^{*\, \prime\, \Lambda}
=
\Lambda^{\Lambda}{}_{\Sigma}\,
\Gamma^{*\, \Sigma}\, ,
\,\,\,\,\,
\Rightarrow
\,\,\,\,\,
\mathcal{B}^{*\, \prime\, \Lambda}
=
\Lambda^{\Lambda}{}_{\Sigma}\, 
\mathcal{B}^{*\Sigma}\, .
\end{equation}

Observe that $\Gamma^{*\, \Lambda}\Gamma_{\Lambda}$ and $\mathcal{B}^{*\,
  \Lambda}\mathcal{B}_{\Lambda}$ are duality invariants.

The simplest equivariant vectors are, up to a complex constant, just equal to the charge vector
$\Gamma^{\Lambda}$. This constant is relevant
because, as we will see, the complex form of the Freudenthal dual of the charge
vector

\begin{equation}
\tilde{\mathcal{Q}}^{M}
=
\left( 
\begin{array}{c}
-2\, \eta^{\Sigma \Lambda}q_{\Lambda}  \\
\\
\tfrac{1}{2}\, \eta_{\Lambda \Sigma} p^{\Lambda}\\
\end{array}
\right)\, ,  
\end{equation}

\noindent
is just $\tilde{\Gamma}^{\Lambda} = -i \Gamma^{\Lambda}$, whence the
phase of the constant corresponds to a global Freudenthal duality
rotation. This immediately implies that the $\mathrm{SU}(1,n)$ invariants $\Gamma^{*\,
  \Lambda}\Gamma_{\Lambda}$ and $\mathcal{B}^{*\,
  \Lambda}\mathcal{B}_{\Lambda}$ are also invariant under Freudenthal
$\mathrm{U}(1)$ duality.  
There may be other equivariant vectors which are functions of the
charges only, but we will not need them. 

We can use the moduli
$Z^{\Lambda}_{\infty}$ in order to construct more equivariant vectors. Again, up to
normalization, the only one we will need is the generic vector given in
Eq.~(\ref{eq:genericmodulidependentequivariantvector}). Multiplying it by
the invariant $\Gamma^{*\, \Lambda}\Gamma_{\Lambda}$ as to give it the
right dimensions for later convenience, we have the
equivariant vector

\begin{equation}
\label{HH}
\Sigma^{\Lambda}
\equiv 
\frac{Z_{\infty}^{*\, \Lambda}}{Z_{\infty}^{*\,
    \Sigma}\Gamma^{*}_{\Sigma}}\ \Gamma^{*\, \Sigma}\Gamma_{\Sigma}\, .
\end{equation}

We will see that in order to find the most general solutions of this
model, it is enough to consider complex linear combinations of the two
equivariant vectors constructed thus far:

\begin{equation}
\label{eq:CPnEqVecComb}
\mathcal{B}^{\Lambda} = \alpha \Gamma^{\Lambda} +\beta \Sigma^{\Lambda}\, ,  
\end{equation}

\noindent
where $\alpha$ and $\beta$ are complex duality invariants (including pure
numbers).

Using this information we can see that in this model (for generic $n$),
in distinction to the axidilaton model, we cannot define a fake charge
$\mathcal{B}^{\Lambda}$ and its associated fake central charge
$\mathcal{Z}(\mathcal{B})$ such that

\begin{equation}
\mathcal{G}^{ij^{*}}\mathcal{D}_{i}\mathcal{Z}
\mathcal{D}_{j^{*}}\mathcal{Z}^{*}
=
|\mathcal{Z}(\mathcal{B})|^{2}
=
e^{\mathcal{K}}|Z^{\Lambda}\Gamma_{\Lambda}|^{2}
-\Gamma^{*\, \Lambda}\Gamma_{\Lambda}\, ,
\end{equation}

\noindent
or such that 

\begin{equation}
V_{\rm bh}(\mathcal{Q}) = V_{\rm bh}(\mathcal{B})\, ,
\end{equation}

\noindent
for arbitrary values of the scalars. This fact has important implications for
the construction of extremal non-supersymmetric solutions as the
first-order equations do not imply the second order ones, which therefore have to be
solved explicitly. In this paper we are going to construct directly the
general non-extremal solutions from which all the extremal ones can be
obtained in the appropriate limits.


\subsection{H-FGK formalism}


The stabilization equations of this model are solved by a linear relation
between $\mathcal{R}_{M}$ and $\mathcal{I}^{M}$, as in the axidilaton case:

\begin{equation}
\mathcal{R}_{M}(\mathcal{I})
=
\mathcal{A}_{MN}\mathcal{I}^{N}\, ,
\hspace{1cm}
\left(\mathcal{A}_{MN}\right)
=
\left( 
\begin{array}{cc}
\tfrac{1}{2}\eta_{\Lambda \Sigma} & 0  \\
& \\
0 & 2 \eta^{\Lambda \Sigma}
 \end{array}
\right)\, ,
\end{equation}

\noindent
which implies that the Freudenthal dual can be expressed as

\begin{equation}
\tilde{H}^{M}
= 
\mathcal{A}^{M}{}_{N}H^N\, ,
\hspace{1cm}
(\mathcal{A}^{M}{}_{N})
=
(\Omega^{PM}\mathcal{A}_{PN})
=
\left( 
\begin{array}{cc}
0 & -2 \eta^{\Lambda \Sigma}  \\
& \\
\tfrac{1}{2}\eta_{\Lambda \Sigma} & 0 \\
\end{array}
\right)\, .
\end{equation}

As in the axidilaton case, $\mathcal{A}_{MN}$ is a symplectic matrix, but, in
contradistinction to that case, $\mathcal{A}^{M}{}_{N}$ is not. In terms of
the complex $H$-variables\footnote{Observe that, in his notation,
  $\mathcal{H}^{\Lambda}\equiv \eta^{\Lambda\Sigma}\mathcal{H}_{\Sigma}$ but
  $H^{\Lambda} \neq \eta^{\Lambda\Sigma}H_{\Sigma}$.}

\begin{equation}
\mathcal{H}_{\Lambda} \equiv
H_{\Lambda}+\tfrac{i}{2}\eta_{\Lambda\Sigma}H^{\Sigma}\, ,  
\end{equation}

\noindent
discrete Freudenthal duality is equivalent to multiplication by a factor of
$-i$.

The Hesse potential reads

\begin{equation}
\mathsf{W}(H)
=
\mathcal{A}_{MN}H^{M} H^{N}
=
\tfrac{1}{2}\eta_{\Lambda \Sigma}H^{\Lambda}H^{\Sigma}
+2\eta^{\Lambda \Sigma}H_{\Lambda}H_{\Sigma}
=
2\mathcal{H}^{*\, \Lambda}\mathcal{H}_{\Lambda}\, ,
\end{equation}

\noindent
and the metric function $e^{-2U}$ and the scalars $Z^{i}$ can be easily
obtained from it as

\begin{equation}
\label{eq:physicalfieldsCPn}
e^{-2U}
=
\mathsf{W}(H)\, ,
\hspace{1cm}
Z^{i}
=
\frac{\tilde{H}^{i}+iH^{i}}{\tilde{H}^{0}+iH^{0}}
=
\frac{H_{i}+\frac{i}{2}H^{i}}{-H_{0}+\frac{i}{2}H^{0}}
=
\frac{\mathcal{H}^{*}_{i}}{\mathcal{H}^{*}_{0}}\, .
\end{equation}

The metric $g_{MN}(H)$ and the potential $V(H)$ have the same structure as in
the axidilaton case when we write them in terms of the matrix
$\mathcal{A}_{MN}$ (which, evidently, is different). Then, the expressions
from Eq.~(\ref{eq:metricaxidilaton}) to
Eq.~(\ref{eq:derivativepotentialaxidil}) are also valid here upon use of the
new matrix $\mathcal{A}_{MN}$. 

The central charge of the model, Eq.~(\ref{eq:centralchargescpn}), takes in
the H-FGK formalism the form

\begin{equation}
\mathcal{Z}(H,\mathcal{Q})
=
-\frac{(H_{0}+\tfrac{i}{2}H^{0})}{\left| H_{0}+\tfrac{i}{2}H^{0} \right|}\
\frac{(\tilde{H}_{M}+iH_{M})\mathcal{Q}^M}{\sqrt{2\mathsf{W}(H)}}.
\end{equation}

It is easy to check that, like in the axidilaton case, this black-hole
potential satisfies

\begin{equation}
V_{\rm bh}(H,\mathcal{Q})
=
\frac{\mathsf{W}(\mathcal{Q})}{\mathsf{W}(H)} \ V_{\rm bh}(\mathcal{Q},H)\, .
\end{equation}


\subsection{Critical points}


Using the complex notation we can write the equation for the critical points
$\mathcal{B}_{\Lambda}$ of the black-hole potential of this model in the form

\begin{equation}
\label{vbhvanish}
\tfrac{i}{2}\mathsf{W}(\mathcal{B})
\left. \partial^{*}_{\Lambda} V_{\rm bh}\right|_{\mathcal{H}=\mathcal{B}}
=
\frac{\mathcal{B}^{\Sigma}\Gamma^{*}_{\Sigma}}{\mathsf{W}(B)}
\left[\mathcal{B}^{*\, \Delta}\Gamma_{\Delta}\mathcal{B}_{\Lambda} 
  -\mathcal{B}^{*\, \Delta}\mathcal{B}_{\Delta}\Gamma_{\Lambda}  \right]
=0\, ,
\end{equation}

\noindent
and can be solved by 

\begin{equation}
\mathcal{B}^{\Sigma}\Gamma^{*}_{\Sigma}=0\, ,
\,\,\,\,\,
\mathrm{or}
\,\,\,\,\,
\mathcal{B}^{*\, \Delta}\Gamma_{\Delta}\mathcal{B}_{\Lambda} 
  -\mathcal{B}^{*\, \Delta}\mathcal{B}_{\Delta}\Gamma_{\Lambda}
=0\, .  
\end{equation}

\noindent
Inserting the general ansatz (\ref{eq:CPnEqVecComb}) into the first condition
we find that it is satisfied for

\begin{equation}
\alpha =-\beta\, , 
\,\,\,\,\,\,
\Rightarrow
\,\,\,\,\,\,
\mathcal{B}^{\Lambda} = \alpha (\Gamma^{\Lambda} -\Sigma^{\Lambda})\, , 
\end{equation}

\noindent
which, up to normalization (which is not fixed in this approach), leaves
us with one arbitrary global phase associated to Freudenthal duality: this is
the moduli-dependent attractor found in Ref.~\cite{Galli:2011fq}.

Inserting our ansatz (\ref{eq:CPnEqVecComb}) into the second condition we get the equation

\begin{equation}
\beta (\alpha^{*} +\beta^{*})   \Gamma^{*\, \Delta}\Gamma_{\Delta}
\Sigma_{\Lambda}
-
\left[
2 \Re\mathfrak{e}(\alpha\beta^{*})
+\frac{|\beta|^{2} \Gamma^{*\, \Sigma}\Gamma_{\Sigma}}{|\mathcal{Z}_{\infty}(\Gamma)|^{2}}
\right] \Gamma^{*\, \Delta}\Gamma_{\Delta}
\Gamma_{\Lambda} =0\, .
\end{equation}

\noindent
The coefficients of the two equivariant vectors must vanish separately, which
can only happen for $\beta=0$, whence $\mathcal{B}^{\Lambda}=\alpha
\Gamma^{\Lambda}$: up to normalization and the Freudenthal duality phase, this is the supersymmetric attractor.


\subsection{Conventional non-extremal solutions}


In this section we are going to show how the knowledge of the
equivariant vectors of the model simplifies the construction of
solutions in the H-FGK formalism.  We are going to see that the most
general solution can be written as

\begin{equation}
\mathcal{H}^{\Lambda}(\tau)=a(\tau)\Gamma^{\Lambda}+b(\tau)\Sigma^{\Lambda}\, ,
\end{equation}

\noindent
where $a(\tau)$ and $b(\tau)$ are two complex, duality-invariant functions
of $\tau$ to be determined. Already, at this stage, we see that this ansatz
reduces dramatically the number of real functions to be found, from $2n+2$ to
just $4$, and all of this without any loss of generality.


First of all, we are going to impose the usual Freudenthal gauge-fixing
condition $\dot{H}^{M}H_{M}=0$ \cite{Galli:2012ji} which in complex notation
takes the form

\begin{equation}
\label{ima}
\Im\mathfrak{m} (\dot{\mathcal{H}}^{*\, \Lambda}\mathcal{H}_{\Lambda})=0\, .
\end{equation}

\noindent
As shown in Ref.~\cite{Galli:2012ji}, assuming this condition, the contraction
of the equations of motion with $H^{M}$ leads to the equation

\begin{equation}
\label{eq:contracted}
\tilde{H}_M\left( \ddot{H}^{M}-r_{0}^{2} H^{M}\right)=0\, ,
\end{equation}

\noindent
which can always be solved by 

\begin{equation}
\label{ans}
\ddot{H}^{M}=r_{0}^{2}H^{M}\, ,
\,\,\,\,\,
\Rightarrow
\,\,\,\,\,
\ddot{\mathcal{H}}^{\Lambda}= r_{0}^{2}\mathcal{H}^{\Lambda}\, .
\end{equation}

\noindent
This is not necessarily the only solution of Eq.~(\ref{eq:contracted}), but as we are going to see
it allows us to solve the rest of the
equations without imposing unnecessary constraints on the physical parameters
of the solution. This equation combined with the equivariant ansatz leads to

\begin{equation}
\label{hh}
\mathcal{H}^{\Lambda}(\tau)
=
\left[c_{1} e^{r_{0}\tau}+c_{3} e^{-r_{0}\tau} \right]\Gamma^{\Lambda}
+\left[c_{2} e^{r_{0}\tau}+c_{4} e^{-r_{0}\tau}  \right]\Sigma^{\Lambda}\, ,
\end{equation}

\noindent
so it only remains to determine the four complex invariants $c_{i}$ $(
i=1,\cdots,4)$ in terms of the charges $\Gamma_{\Lambda}$, the moduli
$Z^{\Lambda}_{\infty}$ and the mass $M$ (or alternatively of the
non-extremality parameter $r_{0}$). 

These four constants can be constrained even further by requiring that the
ansatz gives the right asymptotic behavior for the physical fields in
Eq.~(\ref{eq:physicalfieldsCPn}): requiring that $Z^{\Lambda}_{\infty}=
\mathcal{H}^{*\, \Lambda}_{\infty}/\mathcal{H}^{*\, 0}_{\infty}$ we
get\footnote{
  In the (H-)FGK coordinate system, spatial infinity corresponds to the limit $\tau \rightarrow 0^{-}$.
}

\begin{equation}
\label{eq:c1+c3}
c_{1}+c_{3} = 0\, .  
\end{equation}

\noindent
Asymptotic flatness requires that $\mathcal{H}^{*\,
  \Lambda}_{\infty}\mathcal{H}_{\Lambda, \infty}= \tfrac{1}{2}$ which, upon
use of the above condition, gives

\begin{equation}
\label{e2}
\left|c_{2}+c_{4}\right|^{2}
-\frac{|\mathcal{Z}_{\infty}(\Gamma)|^{2}}{2(\Gamma^{*\, \Lambda}\Gamma_{\Lambda})^{2}}
=
0\, ,  
\end{equation}

\noindent
where $\mathcal{Z}_{\infty}(\Gamma)$ is the central charge at spatial
infinity.  The gauge-fixing condition (\ref{ima}) gives (again, upon use
of Eq.~(\ref{eq:c1+c3}))

\begin{equation}
\label{imi}
\Im \mathfrak{m} \left[c_{3}^{*}(c_{2}+c_{4}) \right]
+\Im \mathfrak{m} \left[c_{2}^{*}c_{4}
\right]\frac{\Gamma^{*\, \Lambda}\Gamma_{\Lambda}}{|\mathcal{Z}_{\infty}(\Gamma)|^{2}}
=0\, .  
\end{equation}

\noindent
Finally, we can still make global Freudenthal duality rotations, which are not
fixed by Eq.~(\ref{ima}): this freedom cannot be used to solve Eq.~(\ref{imi})
but can be used to simplify it by fixing the phase of one of the constants to
a convenient value.

Using the gauge-fixing condition (\ref{ima}), the Hamiltonian constraint
takes the form 

\begin{equation}
\label{hami}
\left[\dot{\mathcal{H}}^{*\, \Lambda}\dot{\mathcal{H}}_{\Lambda}
-\tfrac{1}{2}\Gamma^{*\, \Lambda}\Gamma_{\Lambda}\right]
\mathcal{H}^{*\, \Sigma}\mathcal{H}_{\Sigma}
-2(\dot{\mathcal{H}}^{*\, \Lambda}\mathcal{H}_{\Lambda})^{2}
+\left|\mathcal{H}^{*\, \Lambda}\Gamma_{\Lambda} \right|^{2}
-r_{0}^{2}(\mathcal{H}^{*\, \Lambda}\mathcal{H}_{\Lambda})^{2}
 =  
0\, ,  
\end{equation}

\noindent
and using the gauge-fixing condition plus Eq.~(\ref{ans}) and the Hamiltonian
constraint above, the equations of motion take the form 

\begin{equation}
\label{eqsm}
\mathcal{H}_{\Lambda}^{*}
\left[2(\dot{\mathcal{H}}^{*\, \Sigma}\mathcal{H}_{\Sigma})^{2}
-\left|\mathcal{H}^{*\, \Sigma}\Gamma_{\Sigma}
\right|^{2}\right]
+\Gamma^{*}_{\Lambda}(\mathcal{H}^{*\, \Sigma}\Gamma_{\Sigma})(\mathcal{H}^{*\, \Delta}\mathcal{H}_{\Delta})
-2\dot{\mathcal{H}}_{\Lambda}^{*}
(\dot{\mathcal{H}}^{*\, \Sigma}\mathcal{H}_{\Sigma})(\mathcal{H}^{*\, \Delta}\mathcal{H}_{\Delta})
= 
0\, .
\end{equation}

\noindent
The coefficients of the two equivariant vectors $\Gamma_{\Lambda}$ and
$\Sigma_{\Lambda}$ must vanish independently, which implies that we must solve
the following equations

\begin{align}
\label{eq:eqsm1}
a^{*}
\left[2(\dot{\mathcal{H}}^{*\, \Sigma}\mathcal{H}_{\Sigma})^{2}
-\left|\mathcal{H}^{*\, \Sigma}\Gamma_{\Sigma}
\right|^{2}\right]
+(\mathcal{H}^{*\, \Sigma}\Gamma_{\Sigma})(\mathcal{H}^{*\, \Delta}\mathcal{H}_{\Delta})
-2\dot{a}^{*}
(\dot{\mathcal{H}}^{*\, \Sigma}\mathcal{H}_{\Sigma})(\mathcal{H}^{*\, \Delta}\mathcal{H}_{\Delta})
 =  &
\;0\, ,
\\
 & \nonumber  \\ 
\label{eq:eqsm2}
b^{*}
\left[2(\dot{\mathcal{H}}^{*\, \Sigma}\mathcal{H}_{\Sigma})^{2}
-\left|\mathcal{H}^{*\, \Sigma}\Gamma_{\Sigma}
\right|^{2}\right]
-2\dot{b}^{*}
(\dot{\mathcal{H}}^{*\, \Sigma}\mathcal{H}_{\Sigma})(\mathcal{H}^{*\, \Delta}\mathcal{H}_{\Delta})
 =  &
\;0\, .
\end{align}

\noindent
The coefficients of $b^{*}$ and $\dot{b}^{*}$ in the last equation are real
(on account of the gauge-fixing condition) and this implies that the phases of
$c_{2}$ and $c_{4}$ must be the same up to $\pi$ (the global sign) so that
$\Im\mathfrak{m}(c_{2}^{*}c_{4})=0$ . Then, Eq.~(\ref{imi}) states that the
phase of $c_{3}$ must be the same as that of $c_{2}$ and $c_{4}$, again up to
$\pi$. We know that in the near-horizon limit ({\em i.e.\/} $\tau\rightarrow -\infty$) of
the extremal non-supersymmetric case the phases of $c_{3}$ and $c_{4}$ must differ
by $\pi$ and, since this difference is constant, this must always be
the case. Furthermore, in the extremal non-supersymmetric case
$\mathcal{Z}_{\infty}(\Gamma)=0$ and Eq.~(\ref{e2}) implies that $c_{2}$
and $c_{4}$ must also have opposite global signs. Therefore we find

\begin{equation}
\label{imi2}
\arg(c_{3})= \arg (c_{2}) = \arg(c_{4}) +\pi \equiv \theta \, ,
\end{equation}

\noindent
and, by making use of the global Freudenthal duality freedom

\begin{equation}
\label{e22}
|c_{2}|-|c_{4}|
=
-\frac{|\mathcal{Z}_{\infty}(\Gamma)|}{\sqrt{2}\Gamma^{*\,
    \Lambda}\Gamma_{\Lambda}} \, .
\end{equation}

To simplify the calculations further, we introduce the constant $A$ 

\begin{equation}
\label{c2c4}
|c_{2}|+|c_{4}| 
=
-\frac{|\mathcal{Z}_{\infty}(\Gamma)|}{\sqrt{2}\Gamma^{*\,
    \Lambda}\Gamma_{\Lambda}}A\, , 
\end{equation}

\noindent
which allows us to rewrite Eq.~(\ref{hh}) as

\begin{equation}
\label{hh2}
\mathcal{H}^{\Lambda}(\tau)
=
e^{i\theta}
\left\{ -2|c_{3}| \sinh{r_{0}\tau} \Gamma^{\Lambda}
+
\frac{|\mathcal{Z}_{\infty}(\Gamma)|}{\sqrt{2}\Gamma^{*\,
    \Lambda}\Gamma_{\Lambda}}
\left[(1+A)e^{-r_{0}\tau}+(1-A)e^{r_{0}\tau}  \right]\Sigma^{\Lambda}\right\}\, .
\end{equation}

\noindent
It is now straightforward to solve the equations of motion for the
three constants $\theta$, $A$ and $|c_{3}|$, for which it is
convenient to express the final result using the mass $M$ (defined in
Eq.~(\ref{eq:Mdef}))

\begin{equation}
M
=
r_{0}\left[A+2\sqrt{2}|c_{3}||\mathcal{Z}_{\infty}(\Gamma)|\right]\, .
\end{equation}

\noindent
The final result is

\begin{eqnarray}
\label{c3Atheta}
|c_{3}|
& =  &
\frac{|\mathcal{Z}_{\infty}(\Gamma)|}{2\sqrt{2}Mr_{0}}\, ,
\\
& & \nonumber \\
A
& = &
\frac{M^{2}-|\mathcal{Z}_{\infty}(\Gamma)|^{2}}{Mr_{0}}\, ,
\\
& & \nonumber \\
e^{i\theta}
& = &
\pm\frac{\mathcal{Z}_{\infty}(\Gamma)}{|\mathcal{Z}_{\infty}(\Gamma)|}\, ,
\\
& & \nonumber \\
M^{2}r_{0}^{2} 
& = & 
\left[M^{2}-|\hat{\mathcal{Z}}_{\infty}|^{2} \right]
\left[M^{2}-|\mathcal{Z}_{\infty}(\Gamma)|^{2} \right]\, ,
\end{eqnarray}

\noindent
which is precisely the result obtained in Ref.~\cite{Galli:2011fq}.

We do not expect any other Freudenthal-inequivalent solutions to this model
since the solutions we just found have the maximal number of independent physical
parameters.

\section{The $t^{3}$ model}
\label{sec-t3}


The $t^{3}$-model is characterized by the prepotential

\begin{equation}
\label{eq:t3prepotential}
\mathcal{F}(\mathcal{X})
=
-\tfrac{5}{6}\frac{(\mathcal{X}^{1})^{3}}{\mathcal{X}^{0}}\, .
\end{equation} 

\noindent
In terms of the coordinate $t=\mathcal{X}^{1}/\mathcal{X}^{0}$,
the K\"ahler potential and the scalar-manifold metric are given by

\begin{equation}
\mathcal{K}
=
-3\ln{\Im \mathfrak{m}\, t}
-\ln{\tfrac{20}{3}}\, ,
\hspace{1cm}  
\mathcal{G}_{tt^{*}}
= 
\tfrac{3}{4}\left(\Im\mathfrak{m}\, t\right)^{-2}\, ;
\end{equation}

\noindent
the covariantly holomorphic symplectic section reads

\begin{equation}
\label{eq:symplecticsectiont3}
\mathcal{V}(t,t^{*})
=
e^{\mathcal{K}/2}
\left(
\begin{array}{c}
1 \\
t \\ 
\frac{5}{6}t^{3} \\ 
-\frac{5}{2} t^{2} \\
\end{array}
\right)\, ,
\end{equation}

\noindent
and the central charge, its covariant derivative, the black-hole potential and
its partial derivative read

\begin{eqnarray}
\label{eq:centralcharget3}
\mathcal{Z} 
 & \equiv &
e^{\frac{1}{2}\mathcal{K}}
\hat{\mathcal{Z}}\, , \\
& & \nonumber \\
\label{eq:DtZ}
\mathcal{D}_{t}\mathcal{Z}
& \equiv & 
\tfrac{i}{2}
\frac{e^{\frac{1}{2}\mathcal{K}}}{\Im\mathfrak{m}\, t}
\hat{\mathcal{W}}\, ,
\\
& & \nonumber \\
-V_{\rm bh}
& = & 
e^{\mathcal{K}}
\left[|\hat{\mathcal{Z}}|^{2} +\tfrac{1}{3}|\hat{\mathcal{W}}|^{2}\right]\, ,
\\
& & \nonumber \\
\label{eq:dtV}
-\partial_{t}V_{\rm bh}
& = & 
\tfrac{i}{20}
(\Im \mathfrak{m}\, t)^{-4}
\left[(\hat{\mathcal{W}}^{*})^{2} 
+3 \hat{\mathcal{W}} \hat{\mathcal{Z}}^{*}\right]\, ,
\end{eqnarray} 

\noindent
where we have defined

\begin{eqnarray}
\hat{\mathcal{Z}}
& = & 
\tfrac{5}{6}p^{0} t^{3} -\tfrac{5}{2}p^{1}t^{2} -q_{1} t -q_{0}\, ,
\\
& & \nonumber \\  
\label{eq:hatW}
\hat{\mathcal{W}}
& = & 
\tfrac{5}{2}p^{0}t^{2}t^{*} 
-\tfrac{5}{2}p^{1} t(t+2t^{*})
-q^{1}(2t+t^{*}) -3q^{0}\, .
\end{eqnarray}

\noindent
Observe that all these objects are well defined only iff $\Im\mathfrak{m}\, t >
0$.


\subsection{The global symmetries of the $t^{3}$ model}


The $t^{3}$ model as a theory of $\mathcal{N}=2,d=4$ supergravity is invariant
under global $\mathrm{Sl}(2;\mathbb{R})$ transformations, just like the axidilaton model,
since their K\"ahler metrics are identical up to a numerical factor. The
action of $\mathrm{Sl}(2;\mathbb{R})$ on $t$ is identical to its action on $\lambda$, which was discussed
in Section~\ref{sec-globalsymaxidil}. The transformations of the K\"ahler
potential and covariantly-holomorphic symplectic section
Eqs.~(\ref{eq:Kduality},\ref{eq:Vduality}) are determined by the holomorphic
function $f(t)$ and the $\mathrm{Sp}(4;\mathbb{R})$ matrix $S^{M}{}_{N}$ given by

\begin{eqnarray}
f(t)
& = &   
3\ln{(ct+d)}\, ,
\\
& & \nonumber \\
\label{eq:SMNfort3}
(S^{M}{}_{N})
& = &
\left(
\begin{array}{cccc}
d^{3}               &  3d^{2}c               & \tfrac{6}{5}c^{3}  &
-\tfrac{6}{5}dc^{2}  \\
& & & \\
bd^{2}              & (ad+2bc)d              & \tfrac{6}{5}ac^{2} & -\tfrac{2}{5}(2ad+bc)c \\
& & & \\
\tfrac{5}{6}b^{3}   & \tfrac{5}{2}ab^{2}    & a^{3}              & -a^{2}b  \\
& & & \\
-\tfrac{5}{2}b^{2}d & -\tfrac{5}{2}(2ad+bc)b & -3a^{2}c           & (ad+2bc)a \\ 
\end{array}
\right)\, .
\end{eqnarray}
 
\noindent
In this case the 4-dimensional representation of the generators $T_{m}$
are given by

\begin{equation}
  \begin{array}{cc}
(T_{1}{}^{M}{}_{N})
=
\left(
  \begin{array}{cccc}
3 &   &    &    \\    
  & 1 &    &    \\    
  &   & -3 &    \\    
  &   &    & -1 \\    
  \end{array}
\right)\, ,
\hspace{1.5cm}  
(T_{2}{}^{M}{}_{N})
=
\left(
  \begin{array}{cccc}
   & -3 &   &     \\    
-1 &    &   & 4/5 \\    
   &    &   & 1   \\    
   & 5  & 3 &     \\    
  \end{array}
\right)\, ,
\\
& \\
(T_{3}{}^{M}{}_{N})
=
\left(
  \begin{array}{cccc}
   & -3 &   &     \\    
 1 &    &   & 4/5 \\    
   &    &   & -1  \\    
   & -5 & 3 &     \\    
  \end{array}
\right)\, .
\end{array}
\end{equation}

As in the axidilaton model, the same transformations act on all the symplectic
vectors of the theory and, in particular on $H^{M}$ and
$\mathcal{Q}^{M}$. There are no more symmetries in this formulation of the
model.


\subsubsection{Equivariant vectors of the $t^{3}$ model}


It is not difficult to see that, from the point of view of $\mathrm{Sl}(2;\mathbb{R})$,
the symplectic vectors such as the charge vector $\mathcal{Q}^{M}$ transform
as a quadruplet, i.e.~a fully symmetric 3-index covariant tensor
$\mathcal{Q}_{ijk} = \mathcal{Q}_{(ijk)}$ (in the notation used in
Section~\ref{sec-globalsymaxidil}). The relation between the components of
this tensor and those of the charge vector is

\begin{equation}
\mathcal{Q}_{111}= p^{0}\, ,
\hspace{.5cm}
\mathcal{Q}_{112}=-p^{1}\, ,  
\hspace{.5cm}
\mathcal{Q}_{122}= -\tfrac{2}{5} q_{1}\, ,
\hspace{.5cm}
\mathcal{Q}_{222}= -\tfrac{6}{5} q_{0}\, .
\end{equation}

\noindent
It is useful to observe that the contraction of two quadruplets is related
to the symplectic product by

\begin{equation}
A_{ijk}B^{ijk} = -\tfrac{6}{5}A^{M}B_{M}\, .  
\end{equation}

By definition, any new $\mathrm{Sl}(2;\mathbb{R})$ quadruplet that we construct out of
$t_{\infty}$ and $\mathcal{Q}_{ijk}$ can be transformed according to the above
rules into an equivariant symplectic vector of the $t^{3}$-model. The
$\mathrm{Sl}(2;\mathbb{R})$ index notation makes this construction easy, but, as we are
going to see, insufficient.

In order to construct $\mathrm{Sl}(2;\mathbb{R})$ invariants and other quadruplets it
is useful to define the matrix

\begin{equation}
m^{i}{}_{j} \equiv \mathcal{Q}^{ikl}\mathcal{Q}_{jkl}\, ,
\end{equation}

\noindent
whose components take the values

\begin{equation}
m^{1}{}_{1}= - m^{2}{}_{2} = -\tfrac{2}{5} (p^{1}q_{1} +3p^{0}q_{0})\, ,
\hspace{.5cm}
m^{1}{}_{2}= \tfrac{12}{5}p^{1}q_{0} -\tfrac{8}{25}(q_{1})^{2}\, ,
\hspace{.5cm}
m^{2}{}_{1}=  \tfrac{4}{5}p^{0}q_{1} +2(p^{1})^{2}\, .
\end{equation}

\noindent
The square of this matrix is

\begin{equation}
m^{i}{}_{k}\ m^{k}{}_{j} = -\tfrac{36}{25}\ J_{4}(\mathcal{Q})\ \delta^{i}{}_{j}\, ,  
\end{equation}

\noindent
where, since $\delta^{i}{}_{j}$ is an invariant tensor, the coefficient
$J_{4}(\mathcal{Q})$ must be an invariant of order four in the charges; this
quartic invariant is explicitly given by

\begin{equation}
\label{eq:quarticinvariant}
J_{4}(\mathcal{Q}) \equiv \tfrac{8}{45} p^{0} (q_{1})^{3}
+\tfrac{1}{3}(p^{1}q_{1})^{2} 
-(p^{0}q_{0})^{2} 
-2p^{0}q_{0} p^{1} q_{1} 
-\tfrac{10}{3}(p^{1})^{3} q_{0}\, . 
\end{equation}

This is the only independent invariant that can be constructed from the charge
alone.  We can construct invariants taking traces of powers of $m$ and taking
also the determinant: the traces of odd powers vanish and those of even
powers are proportional to $J_{4}(\mathcal{Q})$. Furthermore, the determinant is also
proportional to $J_{4}(\mathcal{Q})$, {\em i.e.\/}

\begin{equation}
\mathrm{det} (m)
=
\tfrac{36}{25} J_{4}(\mathcal{Q})\, .
\end{equation}

The simplest quadruplet that can be built out of the original one
$\mathcal{Q}_{ijk}$ is

\begin{equation}
\mathcal{Q}_{(ij|l}\ m^{l}{}_{|k)}\, .  
\end{equation}

\noindent
This tensor is necessarily proportional to the Freudenthal dual of
$\mathcal{Q}_{ijk}$ since

\begin{equation}
\mathcal{Q}_{(ij|l}m^{l}{}_{|k)} 
= 
\tfrac{1}{4}\frac{\partial \mathrm{Tr}\, m^{2}}{\partial \mathcal{Q}^{ijk}}
= 
-\tfrac{18}{25}\frac{\partial J_{4}(\mathcal{Q})}{\partial \mathcal{Q}^{ijk}}\, .
\end{equation}

\noindent
Using higher powers of $m$ does not give anything new as

\begin{equation}
\mathcal{Q}_{(i|lm}m^{l}{}_{|j}m^{m}{}_{k)} 
=  
\mathcal{Q}_{(ij|l}m^{l}{}_{m}m^{m}{}_{|k)} 
=
-\tfrac{36}{25}\ J_{4}(\mathcal{Q})\ \mathcal{Q}_{ijk}\, .    
\end{equation}

We must use, therefore, contractions of $\mathcal{Q}_{ijk}$ such that the free
indices are not those of $m^{i}{}_{j}$.  At cubic order in $\mathcal{Q}_{ijk}$
there is only one possibility, which vanishes identically

\begin{equation}
\mathcal{Q}_{(i|lm}\mathcal{Q}_{|j|n}{}^{l}\mathcal{Q}_{|k)}{}^{mn}=0\, ,  
\end{equation}

\noindent
due to the antisymmetry of the symplectic metric $\Omega_{ij}$. At order five
in $\mathcal{Q}_{ijk}$ we can consider

\begin{eqnarray}
\mathcal{Q}_{i,i_{1},i_{2}}
\mathcal{Q}_{j,j_{1},j_{2}}
\mathcal{Q}_{k,k_{1},k_{2}}
\mathcal{Q}^{i_{1},j_{1},k_{1}}
\mathcal{Q}^{i_{2},j_{2},k_{2}} 
& = & 
-\tfrac{36}{25}\ J_{4}(\mathcal{Q})\ \mathcal{Q}_{ijk}\, ,  
\\
& & \nonumber \\
\mathcal{Q}_{(i|mn}\mathcal{Q}_{|j|pq}\mathcal{Q}_{|k)}{}^{mp}m^{nq}
& = &
0\, .
\end{eqnarray}
\noindent
Up to at least order 9 there are no quadruplets other than $\mathcal{Q}_{ijk}$
and its Freudenthal dual that can be constructed by these tensor methods. 

To find more, we have to solve Eq.~(\ref{eq:equivariantvectorequation}). Since
this is a very complicated task, we are going to restrict ourselves to a
generating charge configuration with $p^{0}=q_{1}=0$, \textit{i.e.}

\begin{equation}
(\mathcal{Q}^{M})
=
\left(
  \begin{array}{c}
   0 \\ p^{1} \\ q_{0} \\ 0 \\ 
  \end{array}
\right)\, .
\end{equation}



This subspace is preserved by the $\mathrm{Sl}(2;\mathbb{R})$ transformations with
$b=c=0$ and $d=1/a$ (or equivalently by the infinitesimal transformations
generated by $T_{1}$), to which by analogy we shall refer to as the {\em small group}.  
It is not difficult to see that by acting on this
charge vector with the transformations with appropriate charge-dependent parameters
$b\neq 0\, ,\,\, c\neq0$ (or, equivalently, by the infinitesimal
transformations generated by $T_{2}$ and $T_{3}$) we can generate the complete
generic charge vector with four unrestricted charge components.

It should be clear that if we construct vectors in the subspace
$p^{0}=q_{1}=0$ that are equivariant under the small group, 
then by acting on these vectors with the same transformations that generate the
complete charge vector, we will obtain vectors that are equivariant under
the full duality group, {\em i.e.\/} $\mathrm{Sl}(2;\mathbb{R})$, and which reduce to the former when
we set $p^{0}=q_{1}=0$. Since duality transformations preserve linear independence,
a base for the small-group-equivariant vectors will be transformed into a base of
the duality-group-equivariant vectors; seeing this reasoning we shall refer to a small-group-equivariant vector
as an {\em equivariant-generating vector}.

The equation that these equivariant-generating vectors have to solve is the
restriction of Eq.~(\ref{eq:equivariantvectorequation}) to just $T_{1}$ and allow for no
dependence on $p^{0}$ nor $q_{1}$, {\em i.e.\/}

\begin{equation}
p^{1}\frac{\partial U^{P}}{\partial p^{1}}  
-3 q_{0}\frac{\partial U^{P}}{\partial q_{0}} 
=
\beta^{(P)}U^{(P)}\, ,
\hspace{1cm}
(\beta^{P})
= 
\left(
  \begin{array}{c}
3 \\ 1 \\ -3 \\ -1 \\ 
  \end{array}
\right)\, ,
\end{equation}

\noindent
which is solved by 

\begin{equation}
U^{P}
=
\sum_{i}a_{i}^{(P)} (p^{1})^{\alpha^{(P)}_{i}}
(q_{0})^{\frac{\alpha^{(P)}_{i}-\beta^{(P)}}{3}}\, , 
\end{equation}

\noindent
for arbitrary constants $a_{i}^{P},\alpha^{P}_{i}$ (the parenthesis enclosing
the indices $P$ indicate that they are not summed over and the index $i$ runs
over an arbitrary number of terms). For simplicity, we can choose them to
depend only on $p^{1}$ ($\alpha^{P}= \beta^{P}$) or only on $q_{0}$
($\alpha^{P}_{i}=0$) and take them to have only one term:

\begin{equation}
U^{P} = a^{(P)} (p^{1})^{\beta^{(P)}}\, ,
\hspace{1cm}
U^{P} = a^{(P)} (q_{0})^{-\beta^{(P)}/3}\, .  
\end{equation}

\noindent
To avoid charges with fractional components, we choose the first option and
get a basis of equivariant-generating vectors 

\begin{equation}
U_{\sigma}{}^{P} \sim \delta_{\sigma}{}^{(P)} (p^{1})^{\beta^{(P)}}\, . 
\end{equation}

\noindent
We have found it convenient to normalize these vectors and give them names
$\{R,S,U,V\}$ 

\begin{equation}
R \equiv   
\left(
  \begin{array}{c}
\tfrac{10}{3} (p^{1})^{3} \\ 0 \\ 0 \\ 0 \\ 
  \end{array}
\right)\, ,
\hspace{.2cm}
S \equiv   
\left(
  \begin{array}{c}
0 \\ 0 \\ (\tfrac{10}{3} (p^{1})^{3})^{-1} \\ 0 \\ 
  \end{array}
\right)\, ,
\hspace{.2cm}
U \equiv   
\left(
  \begin{array}{c}
 0 \\ p^{1} \\ 0 \\ 0 \\ 
  \end{array}
\right)\, ,
\hspace{.2cm}
V \equiv   
\left(
  \begin{array}{c}
0 \\ 0 \\ 0 \\ 1/p^{1} \\ 
  \end{array}
\right)\, .
\end{equation}

\noindent
The only non-vanishing symplectic contractions of these four vectors are

\begin{equation}
R_{M}S^{M} = -1\, ,
\hspace{1cm}
U_{M}V^{M}=-1\, ,  
\end{equation}

\noindent
and they satisfy the completeness relation

\begin{equation}
R^{M}S_{N}-S^{M}R_{N} +U^{M}V_{N}  -V^{M}U_{N}
= 
\delta^{M}{}_{N}\, .
\end{equation}

We can decompose any equivariant-generating vector, such as $\mathcal{Q}^{M}$ 
w.r.t.~this basis and the expression will have the same form after acting with the
duality group. For $\mathcal{Q}^{M}$ we find

\begin{equation}
R_{M}\mathcal{Q}^{M} = -\tfrac{10}{3}(p_{1})^{3} q_{0} =
\left. J_{4}(\mathcal{Q}) \right|_{p^{0}=q_{1}=0}\, ,
\hspace{1cm}
V_{M}\mathcal{Q}^{M} =1\, ,
\end{equation}

\noindent
from which we find that in general

\begin{equation}
\mathcal{Q}^{M} = U^{M} -J_{4}(\mathcal{Q}) S^{M}\, .  
\end{equation}

The Freudenthal dual charge vector is (using the results of the next section)
given by

\begin{equation}
\label{eq:Qtildet3}
\tilde{\mathcal{Q}}^{M} 
=
\frac{1}{\mathsf{W}(\mathcal{Q})} R^{M} +\tfrac{3}{4}\mathsf{W}(\mathcal{Q})
V^{M}\, ,
\hspace{1cm}
\mathsf{W}(\mathcal{Q}) = 2\sqrt{J_{4}(\mathcal{Q})}\, .
\end{equation}

As for the moduli-dependent equivariant vectors, we can use the generic
construction in Eq.~(\ref{eq:genericmodulidependentequivariantvector})
replacing $\mathcal{Q}$ with different equivariant vectors.


\subsection{H-FGK formalism}


The stabilization equations can be solved in a completely general way
\cite{Shmakova:1996nz} and the result is summarized by the Hesse
potential which, in terms of the quartic invariant

\begin{equation}
\label{eq:quarticinvariantH}
J_{4}(H) \equiv \tfrac{8}{45} H^{0} (H_{1})^{3}
+\tfrac{1}{3}(H^{1}H_{1})^{2} 
-(H^{0}H_{0})^{2} 
-2H^{0}H_{0} H^{1} H_{1} 
-\tfrac{10}{3}(H^{1})^{3} H_{0}\, , 
\end{equation}

\noindent
can be expressed  as

\begin{equation}
  \mathsf{W}(H)
  =
  2\sqrt{J_{4}(H)}\, .  
\end{equation}

It is convenient to introduce the fully symmetric rank-4
$\mathbb{K}$-tensor \cite{Marrani:2010de,Andrianopoli:2011gy},
implicitly defined by\footnote{In most of what follows, the exact form
  of the $\mathbb{K}$-tensor will be irrelevant. The formulae and
  results obtained will, therefore, be valid for any
  $\mathcal{N}=2,d=4$ theory with Hesse potential of the same generic
  form.}

\begin{equation}
\mathbb{K}_{MNPQ}H^{M}H^{N}H^{P}H^{Q} \equiv J_{4}(H)\, .
\end{equation}

\noindent
Using this tensor, we can write

\begin{eqnarray}
\tilde{H}_{M}
& = &
\frac{\partial_{M} J_{4}}{\mathsf{W}}
=
4\frac{\mathbb{K}_{MNPQ}H^{N}H^{P}H^{Q}}{\mathsf{W}}
\, ,
\\
& & \nonumber \\
\mathcal{M}_{MN}(\mathcal{F})
& = &
-\frac{\partial_{M}\partial_{N} J_{4}}{\mathsf{W}}
+2\frac{\partial_{M} J_{4}\partial_{N} J_{4}}{\mathsf{W}^{3}}
=
-12\frac{\mathbb{K}_{MNPQ}H^{P}H^{Q}}{\mathsf{W}}     
+2\frac{\tilde{H}_{M}\tilde{H}_{N}}{\mathsf{W}}\, ,
\\
& & \nonumber \\
g_{MN}
& = &
24\frac{\mathbb{K}_{MNPQ}H^{P}H^{Q}}{\mathsf{W}^{2}}     
-8\frac{\tilde{H}_{M}\tilde{H}_{N}}{\mathsf{W}^{2}}
-2\frac{H_{M}H_{N}}{\mathsf{W}^{2}}
\, ,
\end{eqnarray}

\noindent
and one can check ({\em e.g.\/} using a symbolic manipulation program) the following properties:

\begin{eqnarray}
J_{4}(\tilde{H})
& = &
J_{4}(H)\, ,
\\
& & \nonumber \\
\mathbb{K}_{MNPQ}\tilde{H}^{N}\tilde{H}^{P}\tilde{H}^{Q}
& = &
-\tfrac{1}{4}\mathsf{W}H_{M}\, ,
\\
& & \nonumber \\
\mathbb{K}_{MNPQ}\tilde{H}^{P}\tilde{H}^{Q}
& = & 
\mathbb{K}_{MNPQ}H^{P}H^{Q}
+\tfrac{1}{6}(H_{M}H_{N}-\tilde{H}_{M}\tilde{H}_{N})\, ,
\\
& & \nonumber \\
\mathbb{K}_{MNPQ}H^{P}\tilde{H}^{Q}
& = & 
-\tfrac{1}{6}H_{(M}\tilde{H}_{N)}\, .
\end{eqnarray}

\noindent
These properties (which hold for any symplectic vector with non-vanishing
quartic invariant which implies the existence of the Freudenthal dual) imply the invariance under
Freudenthal duality of $\mathsf{W}$, $\mathcal{M}_{MN}(\mathcal{F})$ and the
potential $V(H)$; the latter can be rewritten in the manifestly
Freudenthal-duality-invariant form

\begin{equation}
V(H) 
=
-3\mathsf{W}^{-2}
\left\{ 
\mathbb{K}_{MNPQ}\left(H^{P}H^{Q}+\tilde{H}^{P}\tilde{H}^{Q}  \right)
-\tfrac{1}{2}\left(H_{M}H_{N}+\tilde{H}_{M}\tilde{H}_{N} \right)
\right\}\mathcal{Q}^{M}\mathcal{Q}^{N}\, .
\end{equation}

It is, however, not possible to express it in a form manifestly
invariant under the Freudenthal duality transformation of the charge
vector $\mathcal{Q}^{M}\rightarrow \tilde{\mathcal{Q}}^{M}$.

The physical fields are given in terms of the $H$-variables by the usual
expressions

\begin{eqnarray}
e^{-2U}
& = & 
2\mathsf{W}
=
2\sqrt{J_{4}(H)}\, ,
\\
& & \nonumber \\
t 
& = &
\frac{\tilde{H}^{1}+iH^{1}}{\tilde{H}^{0}+iH^{0}}
=
-\frac{3 H^{0}H_{0} + H^{1} H_{1}}{5 (H^{1})^{2} + 2 H^{0}H_{1}}
+i\frac{3\mathsf{W}}{2\left[5 (H^{1})^{2} + 2 H^{0} H_{1}\right]}\, .
\end{eqnarray}


\subsubsection{Very small vectors}


The vectors $R^{M}$ and $S^{M}$ turn out to be very small charge
vectors of this model \cite{Bossard:2012xs,Hristov:2012nu}, owing to
the following properties:

\begin{equation}
\label{eq:verysmallRS}
\mathbb{K}_{MNPQ}R^{P}R^{Q} =  -\tfrac{1}{6}R_{M}R_{N}\, ,
\hspace{1cm}  
\mathbb{K}_{MNPQ}S^{P}S^{Q} =  -\tfrac{1}{6}S_{M}S_{N}\, ,
\end{equation}

\noindent
that leads to (in obvious shorthand notation)

\begin{equation}
\label{eq:verysmallRS2}
\mathbb{K}_{M}R^{3} = \mathbb{K}_{M}S^{3} = 0\, ,
\hspace{1cm}  
J_{4}(R) = J_{4}(S)=0\, .
\end{equation}

On the other hand, the vectors $U^{M}$ and $V^{M}$ are both small vectors

\begin{equation}
\label{eq:verysmallRS3}
J_{4}(U) = J_{4}(V)=0\, .
\end{equation}


\subsection{Critical points}


The complexity of this model forces us to use a symbolic manipulation program
and, further, impose the restriction $p^{0}=q_{1}=0$ on the charges to search
for the critical points of the black-hole potential. Apart from the standard
supersymmetric attractor $B^{M}=\mathcal{Q}^{M}$ we find only one physically
acceptable attractor given by

\begin{equation}
(B^{M})
=
\left(
  \begin{array}{c}
 0 
\\ 
p^{1} 
\\ 
-q_{0}
\\ 
0 
\\   
  \end{array}
\right)\, .
\end{equation}

\noindent
It is an equivariant vector and we can write it in the form 

\begin{equation}
\label{eq:PsAttr}
B^{M} = U^{M}+J_{4}(\mathcal{Q}) S^{M} =\mathcal{Q}^{M}+2J_{4}(\mathcal{Q}) S^{M}\, .
\end{equation}

The quartic invariant for this vector can be computed readily using
Eqs.~(\ref{eq:verysmallRS}--\ref{eq:verysmallRS3}), and 

\begin{equation}
S_{M}\mathcal{Q}^{M}=0\, ,
\hspace{1cm}
\tilde{\mathcal{Q}}_{M}S^{M}=-1/\mathsf{W}(\mathcal{Q})\, ,
\end{equation}

\noindent
and, by Eq.~(\ref{eq:Qtildet3}), it reads

\begin{equation}
\label{eq:Ktensorproperties}
  \begin{array}{rcl}
  J_{4}(B) 
  & =  &
  \mathbb{K} B^{4} = \mathbb{K} [\mathcal{Q} + 2J_{4}(\mathcal{Q}) S]^{4}
  = 
  \mathbb{K}\mathcal{Q}^{4} +8 J_{4}(\mathcal{Q})\mathbb{K}\mathcal{Q}^{3}S
\\
& & \\
& = & 
J_{4}(\mathcal{Q}) 
+2 J_{4}(\mathcal{Q}) \mathsf{W}(\mathcal{Q})\tilde{\mathcal{Q}}_{M}S^{M} 
\\
& & \\
& = & 
-J_{4}(\mathcal{Q})\, .
\end{array}
\end{equation}


\subsection{Conventional extremal solutions}


The supersymmetric solutions of this model are constructed as usual, and we will
focus on the extremal non-supersymmetric ones which are associated to the
attractor $B^{M} = U^{M}+J_{4}(\mathcal{Q}) S^{M}$. For the near-horizon
solutions, the $H^{M}$ take the standard form Eq.~(\ref{eq:NHsolution}) since
Eq.~(\ref{eq:conditionNHsolution}) is satisfied. Now we must investigate
whether we can add constant terms $A^{M}$ to these harmonic functions
satisfying only the normalization condition $\mathsf{W}(A)=1$ and the
constraint $B^{M}A_{M}=0$, which is equivalent, at the infinitesimal level, to
investigating the space of solutions to
Eq.~(\ref{eq:conditioninfinitesimalextension}). For simplicity, we work with a
generating charge configuration with $p^{0}=q_{1}=0$. We find for the
non-supersymmetric attractor

\begin{equation}
  (\mathfrak{M}_{MN})
=
\tfrac{1}{2} 
\left(
  \begin{array}{cccc}
  \tfrac{21}{20}\frac{q_{0}}{(p^{1})^{3}} & 0 & 0 & -\tfrac{3}{20}\frac{1}{(p^{1})^{2}} \\
  0 & 0 & 0 & 0 \\
  0 & 0 & 0 & 0 \\
-\tfrac{3}{20}\frac{1}{(p^{1})^{2}} & 0 & 0 & \tfrac{1}{4}
\frac{1}{p^{1}q_{0}} \\
  \end{array}
\right)\, ,
\end{equation}

\noindent
whose rank is 2. The solutions to
Eq.~(\ref{eq:conditioninfinitesimalextension}) have the form
$(\varepsilon^{M})= \left(
\begin{smallmatrix}
0 \\ \varepsilon^{1} \\ \varepsilon_{0} \\ 0 \\  
\end{smallmatrix}
\right)$ and satisfy $B^{M}\varepsilon_{M}=0$ but we still have to
impose the normalization condition $\mathsf{W}(A)=1$ on the two
non-vanishing components, which leaves us with only one independent
solution that can only describe one independent real moduli; this
modulus turns out to be $\Im\mathfrak{m}(t_{\infty})$.  It can be
shown that the solution takes the form \cite{Galli:2012jh}

\begin{equation}
\label{sol:Effe}
\left(H^{M}\right)
=
\left(
  \begin{array}{c}
 0 
\\ 
\\
s^{1}
\left\{
\sqrt{\frac{3}{10\Im\mathfrak{m}\, t_{\infty}}} 
-\frac{1}{\sqrt{2}}|p^{1}|\tau 
\right\} 
\\ 
\\
-s_{0}\left\{ \sqrt{\frac{5 (\Im\mathfrak{m}\, t_{\infty})^{3}}{24}}  
-\tfrac{1}{\sqrt{2}}|q_{0}|\tau\right\} 
\\ 
\\
0 
\\   
  \end{array}
\right)\, ,  
\end{equation}

\noindent
where we have defined

\begin{equation}
s^{M} \equiv \operatorname{sgn}(\mathcal{Q}^{M})\, ,  
\end{equation}

\noindent
and where we have to require $s^{1}=s_{0}$ for the solution to be regular.

Having $\Re\mathfrak{e}\, t_{\infty}=0$ poses a very important problem because
even though the charge vector with $p^{0}=q_{1}$ can generate via
$\mathrm{Sl}(2;\mathbb{R})$ duality transformations a complete charge vector with four
independent charges, it cannot  at the same time generate an independent $\Re\mathfrak{e}\,
t_{\infty}\neq 0$. In other words, this solution is not a
\textit{generating solution}; its orbit under $\mathrm{Sl}(2;\mathbb{R})$ rotations
will not fully cover the space of parameters.  A necessary and sufficient
condition for a solution to be generating is that all the $\mathrm{Sl}(2;\mathbb{R})$
invariants of the theory are independent when evaluated on the charges and
moduli of that solution \cite{Bertolini:1998mt,Bertolini:1999je}. As we show
in detail in Appendix~\ref{sec-2charget3}, the solution (\ref{sol:Effe}) does not satisfy this
condition.

In order to have a generating solution for the class of extremal
non-supersymmetric black-hole solutions associated to the attractor $B^{M} =
U^{M}+J_{4}(\mathcal{Q}) S^{M}$, we need to add $\Re\mathfrak{e}\,
t_{\infty}\neq 0$ to the solution and it should be clear that this cannot be
done if we make a conventional, {\em i.e.\/} harmonic, ansatz: the $H^{M}$ must contain
anharmonic terms. 

For future use, it is useful to have symplectic-covariant expressions for the
constraints on $A^{M}$ imposed by the equations of motion for a harmonic
ansatz:

\begin{equation}
\label{eq:AMextraconstraints}
A_{M}U^{M} =0\, ,
\hspace{1cm}
A_{M}S^{M}=0\, .
\end{equation}

\noindent
$A_{M}B^{M}=0$ only imposes the weaker condition
$A_{M}(U^{M}+J_{4}(\mathcal{Q})S^{M})=0$. The above constraints imply that
$A^{M}$ has to take the form

\begin{equation}
A^{M} = aU^{M}+bS^{M}\, ,  
\end{equation}

\noindent 
for some invariant coefficients $a$ and $b$, and it cannot contain terms
proportional to the vectors $R^{M}$ and $V^{M}$.


\subsection{Unconventional extremal solutions}
\label{eq:t3unconventional}


The missing free parameter must be added to the above solution by adding
anharmonic terms to the harmonic ansatz: let us don the harmonic
functions of the undeformed solution with hats, so that

\begin{equation}
\hat{H}^{M} = A^{M}- \tfrac{1}{\sqrt{2}}B^{M}\tau\, ,  
\end{equation}

\noindent
where $B^{M}$ is given by the attractor (\ref{eq:PsAttr}) and $A^{M}$ satisfies the
constraints Eqs.~(\ref{eq:AMextraconstraints}) but is otherwise arbitrary (up to
asymptotic flatness normalization). Observe that this implies that 

\begin{equation}
\label{eq:hatHextraconstraints}
\hat{H}_{M}U^{M} =  \hat{H}_{M}S^{M} =  0\, ,
\,\,\,\,\,
\Rightarrow
\,\,\,\,\,
\hat{H} = a(\tau) U^{M} + b(\tau) S^{M}\, ,
\end{equation}

\noindent
where $a(\tau)$ and $b(\tau)$ are duality-invariant harmonic functions of
$\tau$. Terms proportional to $R^{M}$ and $V^{M}$ are excluded if the
coefficients are harmonic functions; a term proportional to $V^{M}$ can
always be eliminated by a local Freudenthal duality transformation, whence
we expect that it is enough to add a (necessarily anharmonic) term
proportional to $R^{M}$. It turns out that such a solution
\cite{Galli:2012jh}\footnote{This solution can be obtained by truncation from
  the STU-model solution in Ref.~\cite{Gimon:2009gk} and is also a particular
  case of the general extremal non-supersymmetric solutions of cubic models of
  Ref.~\cite{Bossard:2012xs}. It has also been obtained by using integrability
  methods in the action that one obtains in the approach of
  Ref.~\cite{Breitenlohner:1987dg} (see also \cite{Chemissany:2010zp}): its
  derivation can be found in Section~9.4 (page 76) of
  Ref.~\cite{Fre:2011uy}. The solution belongs to the orbit
  $\mathcal{O}^{3}_{22}$ in the classification of Ref.~\cite{Fre:2012im} (see
  Table~2 of that reference).} has the form\footnote{This definition is not
  recursive because $R_{N}H^{N}=R_{N}\hat{H}^{N}$.}

\begin{equation}
\label{eq:t3nonBPSgeneralsolution}
H^{M} = \hat{H}^{M} -\frac{\chi R^{M}}{R_{N}H^{N}} \, ,
\end{equation}

\noindent
where $\chi$ is another independent parameter, like $A^{M}$. The values of
$\chi$ and $A^{M}$ are determined by requiring that the physical fields have
the right asymptotic behavior at spatial infinity ($e^{-2U}\rightarrow 1\,
,\,\,\, t\rightarrow t_{\infty}$ when $\tau\rightarrow 0^{-}$) as
follows: first of all, observe that as a consequence of
Eq.~(\ref{eq:hatHextraconstraints}) the property

\begin{equation}
\label{eq:HU=0}
H_{M}U^{M} = 0 \, , 
\end{equation}

\noindent
is satisfied everywhere and in particular at spatial infinity where

\begin{equation}
\label{eq:Hinf}
H^{M} \stackrel{\tau\rightarrow 0^{-}}{\longrightarrow} H^{M}_{\infty} =
A^{M} -\frac{\chi R^{M}}{R_{N}A^{N}}\, .
\end{equation}

\noindent
Then, using the definition of $H^{M}=\mathcal{I}^{M}$, Eq.~(\ref{eq:RandIdef}),
in Eq.~(\ref{eq:HU=0}) plus Eq.~(\ref{eq:Xdef}) at spatial infinity we find

\begin{equation}
0
=
H_{M\, \infty} U^{M}
= 
\Im\mathfrak{m}\left(\frac{\mathcal{V}_{M\,\infty}}{X_{\infty}}\right) U^{M}
=
\Im\mathfrak{m}\left(\frac{\mathcal{Z}_{\,\infty}(U)}{X_{\infty}}\right)
=
\sqrt{2}
\Im\mathfrak{m}\left(\frac{\mathcal{Z}_{\,\infty}(U)}{e^{i\alpha_{\infty}}}\right)\,
.
\end{equation}

\noindent
This implies that

\begin{equation}
e^{i\alpha_{\infty}}
=
\pm \frac{\mathcal{Z}_{\infty}(U)}{|\mathcal{Z}_{\infty}(U)|}\, ,  
\end{equation}

\noindent
which can be used again in the definition of $H^{M}=\mathcal{I}^{M}$ to give

\begin{equation}
\label{eq:HMinf}
H^{M}_{\infty}     
= 
\pm \sqrt{2}
\Im\mathfrak{m}\left(\frac{\mathcal{V}^{M}_{\infty}}{\mathcal{Z}_{\infty}(U)}\right)
|\mathcal{Z}_{\infty}(U)|\, .
\end{equation}

\noindent
To determine the overall sign we will demand that the functions $H^{M}(\tau)$
never vanish for $\tau \in [-\infty,0)$, a condition that is usually related
to the positivity of the mass. Contracting the above result with $S^{M}$ and
using Eq.~(\ref{eq:hatHextraconstraints}) we get

\begin{equation}
\label{eq:chiRA}
\frac{\chi}{R_{N}A^{N}} 
=   
\pm \sqrt{2}
\Im\mathfrak{m}\left(\frac{\mathcal{Z}_{\infty}(S)}{\mathcal{Z}_{\infty}(U)}\right)
|\mathcal{Z}_{\infty}(U)|\, ,
\end{equation}

\noindent
which, after substitution in Eq.~(\ref{eq:Hinf}) gives the value of the
constants $A^{M}$, satisfying Eqs.~(\ref{eq:AMextraconstraints}), as an
equivariant symplectic vector, function of the physical parameters of the
solution

\begin{equation}
A^{M} 
= 
\pm \sqrt{2}
(\delta^{M}{}_{N} -R^{M}S_{N})  
\Im\mathfrak{m}\left(\frac{\mathcal{V}^{M}_{\infty}}{\mathcal{Z}_{\infty}(U)}\right)
|\mathcal{Z}_{\infty}(U)|\, .
\end{equation}

With this information we can compute $R_{N}A^{N}$ to find, from
Eq.~(\ref{eq:chiRA}) the value of the invariant parameter $\chi$ as a function
of the physical parameters of the solution\footnote{In terms of the invariants
  $i_{1},\cdots, i_{5}$ of the theory given in
  Eqs.~(\ref{eq:invariant1})-(\ref{eq:invariant5}) 
\begin{equation}
\chi 
=
\tfrac{1}{4}(-J_{4}(\mathcal{Q}))^{-1/6}
\left\{
\left(
i_{1} + i_{2} - \frac{(i_{1} - i_{2}\,/3)^3}{J_{4}(\mathcal{Q})} - \frac{4\,
  i_{3}}{\sqrt{-J_{4}(\mathcal{Q})}}
\right)^{1/3} 
- \left(i_{1} + i_{2} - \frac{(i_{1} - i_{2}\,/3)^3}{J_{4}(\mathcal{Q})}+ \frac{4\, i_{3}}{\sqrt{-J_{4}(\mathcal{Q})}}\right)^{1/3}
\right\}\,.
\end{equation}
}

\begin{equation}
\chi  
=
-2 
\Im\mathfrak{m}\left(\frac{\mathcal{Z}_{\infty}(R)}{\mathcal{Z}_{\infty}(U)}\right)
\Im\mathfrak{m}\left(\frac{\mathcal{Z}_{\infty}(S)}{\mathcal{Z}_{\infty}(U)}\right)
|\mathcal{Z}_{\infty}(U)|^{2}\, .
\end{equation}

For $p^{0}=q_{1}=0$, the solution takes the explicit (but not manifestly
equivariant) form

\begin{equation}
\left(H^{M} \right)
=
\left(
  \begin{array}{c}
 -\tfrac{1}{2} \frac{\Re\mathfrak{e}\, t_{\infty} }{\Im\mathfrak{m}\, t_{\infty}}
{\displaystyle\frac{1}{H_{0}}}\, ,
\\ 
\\
s^{1}
\left\{
\sqrt{\frac{3}{10\Im\mathfrak{m}\, t_{\infty}}} 
-\frac{1}{\sqrt{2}}|p^{1}|\tau 
\right\} 
\\ 
\\
-s_{0} 
\left(\frac{|t_{\infty}|}{\Im\mathfrak{m}\, t_{\infty}}\right)^{2}
\left\{ \sqrt{\frac{5 \Im\mathfrak{m}\, t_{\infty}}{24}}  
-\tfrac{1}{\sqrt{2}}|q_{0}|\tau\right\} 
\\ 
\\
0 
\\   
  \end{array}
\right)\, .  
\end{equation}

The mass of this solution can be computed using the general formula
Eq.~(\ref{eq:Mdef}). From the definition of $\tilde{H}_{M}$ we have

\begin{equation}
\tilde{H}_{M} (0)
= 
\pm \sqrt{2}\Re\mathfrak{e}\left(\frac{\mathcal{V}_{\infty\, M}}{\mathcal{Z}_{\infty}(U)} \right) |\mathcal{Z}_{\infty}(U)|\, ,
\end{equation}

\noindent
and

\begin{equation}
\dot{H}^{M}(0) 
= 
-\tfrac{1}{\sqrt{2}}
\left[B^{M} -\frac{\chi J_{4}(\mathcal{Q})}{(RA)^{2}} R^{M}\right]\, ,
\end{equation}

\noindent
from which we get the covariant expression

\begin{equation}
M
=
\pm |\mathcal{Z}_{\infty}(U)| 
\left\{ 
1 -\tfrac{1}{3} J_{4}(\mathcal{Q}) 
\Im\mathfrak{m} \left(\frac{\mathcal{Z}_{\infty}(V)}{\mathcal{Z}_{\infty}(U)}\right)
\left[
\Im\mathfrak{m} \left(\frac{\mathcal{Z}_{\infty}(R)}{\mathcal{Z}_{\infty}(U)}\right)
\right]^{-1}
\right\}\, .
\end{equation}

\noindent
This last expression reduces for $p^{0}=q_{1}=0$ (selecting the upper sign in
Eq.~(\ref{eq:HMinf})) to

\begin{equation}
M = e^{\mathcal{K}_{\infty}/2} 
\left(
|q_{0}| + \tfrac{5}{2}|t_{\infty}|^{2} |p^{1}|
\right)\, .  
\end{equation}

\noindent
Observe that the value of the mass differs from the absolute value of the
associated fake central charge $B^{M}$:

\begin{equation}
M\neq |\mathcal{Z}(\phi_{\infty},B)|\, . 
\end{equation}

The above result should be compared to the mass of the supersymmetric
black hole which is given by the standard formula
$M=|\mathcal{Z}_{\infty}(\mathcal{Q})|$ and reduces for
$p^{0}=q_{1}=0$ to\footnote{We have used that $p^{1}q_{0}>0$ for the
  non-supersymmetric case and $p^{1}q_{0}<0$ for the supersymmetric
  one.} the following expression,

\begin{equation}
M 
=
e^{\mathcal{K}_{\infty}/2} 
\sqrt{
\left[|q_{0}| - \tfrac{5}{2}(\Re\mathfrak{e}t_{\infty})^{2} |p^{1}|\right]^{2} 
+\tfrac{25}{4}(\Im\mathfrak{m} t_{\infty})^{4} |p^{1}|^{2} 
+5 (\Im\mathfrak{m}  t_{\infty})^{2}|q_{0}p^{1}|
}\, ,   
\end{equation}

\noindent
which can be rewritten in the equivalent form 

\begin{equation}
M 
=
e^{\mathcal{K}_{\infty}/2} 
\sqrt{
\left[|q_{0}| +\tfrac{5}{2}|t_{\infty}|^{2} |p^{1}|\right]^{2} 
-10 (\Re\mathfrak{e} t_{\infty})^{2}|q_{0}p^{1}|
}\, ,   
\end{equation}

\noindent
which shows that the mass of the supersymmetric black hole is always
smaller than the mass of the non-supersymmetric one with charges of
equal absolute value.

The entropy is given by the square of the fake central charge at the horizon

\begin{equation}
S = \pi |\mathcal{Z}(\phi_{h},B)|^{2} = \pi \mathsf{W}(B)/2 = \pi
\sqrt{-J_{4}(\mathcal{Q})}\, .   
\end{equation}

As discussed in Section~\ref{sec-firstorder}, an interesting
characteristic of the unconventional solutions is that, in distinction
to what happens for the conventional ones, the flow of the black-hole
metric function $e^{-U}$ from infinity to the horizon is not governed
by a simple fake central charge $\mathcal{Z}(\phi ,B)$ since the
near-horizon limit of the metric is related to $\mathcal{Z}(\phi_{\rm
  h},B)$ but the spacelike infinity limit is not related to
$\mathcal{Z}(\phi_{\infty},B)$. The first-order flow equations for
these black holes can be written in terms of a superpotential
$W(\phi,B)$ or, equivalently, in terms of the ``fake central charge''
$\mathcal{Z}(\phi,\sqrt{2}\mathfrak{D}H)$ defined in
Section~\ref{sec-firstorder}. 

It is possible to prove analytically that the general configuration
Eq.~(\ref{eq:t3nonBPSgeneralsolution}) solves the equations of motion
by using the duality-invariant properties of the equivariant vectors
$A^{M}$, $B^{M}$ and $R^{M}$ that appear in its definition (that is:
not reducing the equations to the $p^{0}=q_{1}$ case) and the
properties of the $\mathbb{K}$-tensor of this model, see
Eqs.~(\ref{eq:Ktensorproperties}). As an intermediate step, we derive
the following relations, which are valid only for the $H^{M}s$ of our
ansatz:

\begin{eqnarray}
\mathbb{K}_{MN}\hat{H}^{2}
& = & 
\tfrac{1}{2}(VH)^{2} R_{(M}V_{N)}  
+\tfrac{1}{2}(VH)(RH) V_{M}V_{N}
+\tfrac{1}{18}(VH)^{2} U_{M}U_{N}  
\nonumber \\
& & \nonumber \\
& & 
-\tfrac{1}{3}(VH)(RH) U_{(M}S_{N)}  
-\tfrac{1}{6}(RH)^{2} S_{M}S_{N}\, ,  
\\
& & \nonumber \\
\mathbb{K}_{MN}\hat{H}\mathcal{Q}
& = & 
\tfrac{1}{2}(VH) R_{(M}V_{N)}  
+\tfrac{1}{4}[J_{4}(\mathcal{Q})(VH)+(RH)] V_{M}V_{N}
+\tfrac{1}{18}(VH) U_{M}U_{N}  
\nonumber \\
& & \nonumber \\
& & 
-\tfrac{1}{6}[J_{4}(\mathcal{Q})(VH)+(RH)] U_{(M}S_{N)}  
-\tfrac{1}{6}J_{4}(\mathcal{Q})(RH) S_{M}S_{N}\, ,  
\\
& & \nonumber \\
\mathbb{K}_{MN}\hat{H}R
& = & 
-\tfrac{1}{3}(RH) R_{(M}S_{N)}  
-\tfrac{1}{6}(RH) U_{(M}V_{N)}
-\tfrac{1}{6}(VH) R_{(M}U_{N)}\, .  
\end{eqnarray}

Using these identities it is easy to show, for instance, that

\begin{equation}
J_{4}(H) = J_{4}(\hat{H}) -\chi^{2}\, ,
\hspace{1cm}  
J_{4}(\hat{H}) = (VH)^{3}\ (RH)\, .
\end{equation}


\section{Conclusions}
\label{sec-conclusions}


In this paper we have shown how the equivariance of the $H$ variables
under duality transformations translates into equivariance of the
constant symplectic vectors that occur in their explicit
expressions. Using the H-FGK formalism we have studied under what
conditions the extremal solutions associated to a given attractor can
be described, for all values of the charges and moduli, by harmonic
$H$s alone and when it is necessary to add anharmonic terms to
them. We have called these two kinds of solutions conventional,
respectively unconventional.

As mentioned in the introduction, it is not known how unconventional
extremal solutions (which are necessarily non-supersymmetric, since we
know that all the supersymmetric ones are conventional) can be
deformed into non-extremal solutions, with non-zero temperature but
the same values of the charges and moduli. The H-FGK formalism and the
use of equivariant vectors can help us to solve this problem and, as a
first step, we have shown how to apply these methods to well-known
examples of theories with conventional and unconventional solutions.

In the case of the unconventional extremal solutions of the
$t^{3}$-model we have shown, first of all, how the criterion found in
Section~\ref{sec-conventional} indicates the need for anharmonic terms
and which equivariant vectors these terms should depend on. We have
then described the solution entirely in terms of these objects and we
have computed the general form of the mass and the entropy. The second
has a well-known form in terms of the near-horizon limit
$\mathcal{Z}(\phi_{\rm h},B)$ of a \textit{fake central charge},
$\mathcal{Z}(\phi,B)$, constructed from what we have called (in the
context of the H-FGK formalism) attractor $B^{M}$. The mass instead is
not given by the spacelike infinity limit of this fake central charge
$M=|\mathcal{Z}(\phi_{\infty},B)|$ but rather by the spacelike
infinity of a different one $\mathcal{Z}(\phi,E)$ with $E^M\neq
B^M$. The first-order flow equations that govern the system (which
have been given in Refs.~\cite{Galli:2010mg,Bossard:2012xs}) are
written in term of non-standard fake central charge
$\mathcal{Z}(\phi,\sqrt{2}\mathcal{D}H)$ whose second argument is
$\tau$-dependent and correctly interpolates between $B^{M}$ (on the
horizon) and $E^{M}$ (at spacelike infinity). 

The behavior of the metric function in the unconventional solutions
gets modified in the asymptotic region but remains unchanged in the
near-horizon region, where it is still governed by the attractor
mechanism. This behavior is reminiscent, but opposite, to that of the
colored non-Abelian supersymmetric black holes of
Refs.~\cite{Meessen:2008kb} in which the near-horizon geometry is
modified by the non-Abelian effects while the asymptotic one is
unchanged by them.

The formalism and the methods presented in this paper can be applied
to the problem of finding the non-extremal generalization of the
unconventional solutions studied in this paper. Work in this direction
is in progress.


\section*{Acknowledgments}


TO would like to thank Pietro Fr\'e, Alessio Marrani, Thomas Van Riet,
Jan Perz and Renata Kallosh for very useful conversations and the
String Theory Group of the K.U.~Leuven and the INFN section of the
University of Padova for its hospitality.  PM and PG would like to
thank the Instituto de F\'{\i}sica Te\'orica its hospitality.  This
work has been supported in part by the Spanish Ministry of Science and
Education grant FPA2012-35043-C02-01, the Comunidad de Madrid grant
HEPHACOS S2009ESP-1473 and the Spanish Consolider-Ingenio 2010 program
CPAN CSD2007-00042.  The work of PB has been supported by the
JAE-predoc grant JAEPre 2011 00452.  The work of PG has been supported
in part by grants FIS2008-06078-C03-02 and FPA2008-03811-E/INFN of
Ministerio de Ciencia e Innovaci\'on (Spain) and ACOMP/2010/213 from
Generalitat Valenciana. The work of PM has been supported by the
Ram\'on y Cajal fellowship RYC-2009-05014. TO wishes to thank
M.M.~Fern\'andez for her permanent support.

\appendix

\section{Generating new solutions via duality}
\label{app:generating}

As mentioned in Section~\ref{eq:t3unconventional}, a necessary and
sufficient condition for a solution to be generating is that all the
$\mathrm{Sl}(2;\mathbb{R})$ invariants of the theory are independent
when evaluated on the charges and moduli of that solution
\cite{Cvetic:1996zq,Andrianopoli:1997wi,Bertolini:1998mt,Bertolini:1999je}. In
this appendix we are going to study whether or not and why the
solution considered in that section is a generating one. We start by
stating some general properties which we, then, apply to the (toy)
axidilaton model and then to the $t^{3}$ model.

There are in general 5 independent invariants that characterize each
$\mathcal{N}=2$ symmetric supergravity model. They are
\cite{Cerchiai:2009pi}:

\begin{eqnarray}
\label{eq:invariant1}
i_{1} 
& = &
|\mathcal{Z}|^{2}\, ,
\\
& & \nonumber \\
i_{2} 
&  = &
\mathcal{G}^{ij^{*}} \mathcal{Z}_{i}\mathcal{Z}^{*}_{j^{*}}\,,
\\
& & \nonumber \\
i_{3} 
& = & 
-\tfrac{1}{3} \Re\mathfrak{e} 
\left[\mathcal{Z}\mathcal{N}_{3}(\mathcal{Z}^{*})\right]\, ,
\\
& & \nonumber \\
i_{4}
& = & 
\tfrac{1}{3}\Im\mathfrak{m}
\left[\mathcal{Z}\mathcal{N}_{3}(\mathcal{Z}^{*})\right]\, ,
\\
& & \nonumber \\
\label{eq:invariant5}
i_{5}
& = &
\mathcal{G}^{ij^{*}} 
\mathcal{C}_{ijk}\mathcal{C}^{*}_{i^{*}j^{*}k^{*}}
\mathcal{G}^{jl^{*}}\mathcal{G}^{km^{*}}\mathcal{G}^{j^{*}l}
\mathcal{G}^{k^{*}m}
\mathcal{Z}^{*}_{l^{*}}\mathcal{Z}^{*}_{m^{*}}
\mathcal{Z}_{l}\mathcal{Z}_{m}\, ,
\end{eqnarray}

\noindent
where $\mathcal{Z}$ is the central charge, $\mathcal{G}^{ij^{*}}$ the
inverse K\"ahler metric,

\begin{equation}
\mathcal{Z}_{i}\equiv \mathcal{D}_{i} \mathcal{Z}\, ,
\end{equation}

\noindent
are the ``matter'' central charges, 

\begin{equation}
\mathcal{C}_{ijk} 
\equiv 
\mathcal{D}_{i}\mathcal{V}_{M} \mathcal{D}_{j}\mathcal{D}_{k}\mathcal{V}^{M}\, ,  
\end{equation}
 
\noindent
and 

\begin{equation}
\mathcal{N}_{3}(\mathcal{Z}^{*})
\equiv
\mathcal{C}_{ijk} 
\mathcal{G}^{il^{*}}\mathcal{G}^{jm^{*}}\mathcal{G}^{kn^{*}}
\mathcal{Z}^{*}_{l^{*}}\mathcal{Z}^{*}_{m^{*}}\mathcal{Z}^{*}_{n^{*}}\, .
\end{equation}

\noindent
All these invariants are function of the charges and the scalars but
their combination

\begin{equation}
 J_{4}(\mathcal{Q})=(i_{1}-i_{2})^{2}+4i_{4}-i_{5}\, ,
\end{equation}

\noindent
depends quartically on the charges only. Sometimes it is advantageous
to work with $J_{4}(\mathcal{Q})$ instead of $i_{5}$.


\subsection{2-charge generating solutions of the axidilaton model}


The minimal number of non-vanishing charges that are necessary for an
extremal, supersymmetric\footnote{The discussion can also be held for
  the non-supersymmetric solutions to this model, reaching the same
  conclusions.}, black hole of axidilaton theory to be regular is
two. Taking into account the form of the Hesse potential
Eq.~(\ref{eq:Hesseaxidilaton}) and of the axidilaton
Eq.~(\ref{eq:axidilaton}), it is easy to see that there are only two
possible non-singular 2-charge configurations, namely
$(p^{0},p^{1},0,0)^{T}$ and $(0,0,q_{0},q_{1})^{T}$.

In this model, the tensor $\mathcal{C}_{ijk}$ vanishes identically,
and so does $\mathcal{N}_{3}(\mathcal{Z}^{*})$ and the invariants
$i_{3},i_{4},i_{5}$. The model is characterized by the two invariants
$i_{1}$ and $i_{2}$, which are, respectively, the squares of the
absolute values of the true and fake central charges at infinity

\begin{equation}
i_{1}=|\mathcal{Z}(\lambda_{\infty},\mathcal{Q})|^{2}\, ,
\hspace{1cm}
i_{2}=|\hat{\mathcal{Z}}(\lambda_{\infty},\mathcal{Q})|^{2}\, ,  
\end{equation}


\noindent
and both are independent for any 2-charge solution (for
$\Re\mathfrak{e}\, \lambda_{\infty}=0$ or not) and, in principle, it
should be a generating solution.  However, depending on our choice of
harmonic functions, the regular solutions with two charges may have a
vanishing $\Re\mathfrak{e}\, \lambda_{\infty}$ and the subgroup of
$\mathrm{Sl}(2;\mathbb{R})$ that generates a non-vanishing
$\Re\mathfrak{e}\, \lambda_{\infty}$, which
consists of matrices of the form $\left(\begin{smallmatrix} 1 & \beta \\ 0 & 1 \\
  \end{smallmatrix} \right)$ do not leave invariant the 2-charge
configurations. Therefore, the $\mathrm{Sl}(2;\mathbb{R})$ orbit of the regular
2-charge configurations may not cover the full parameter space.

It is interesting to see how the impossibility of generating a solution
containing the maximal number of independent parameters arises in
practice in this simple case, starting from a configuration
characterized by the charges $(0,0,\hat{q}_{0},\hat{q}_{1})^{T}$
and the moduli $\hat{\lambda}_{\infty}= i \Im\mathfrak{m}\,
\hat{\lambda}_{\infty}$ (we reserve the unhatted symbols for the
final charges and moduli). This solution is determined by two harmonic
functions:

\begin{equation}
(\hat{H}^{M}) 
=
\left(
  \begin{array}{c}
  0 
\\ 
\\  
0
\\
\\
\frac{s}{\sqrt{2}} \left\{ (\Im\mathfrak{m}\,
\hat{\lambda}_{\infty})^{1/2} - |\hat{q}_{0}| \tau \right\} 
\\
\\
\frac{s}{\sqrt{2}} \left\{ (\Im\mathfrak{m}\,
\hat{\lambda}_{\infty})^{-1/2} - |\hat{q}_{1}| \tau \right\}
\\
  \end{array}
\right)\, ,  
\end{equation}

\noindent
where 

\begin{equation}
s \equiv \operatorname{sgn}(\hat{q}_{0}) 
= 
\operatorname{sgn}(\hat{q}_{1})\, .
\end{equation}

The $\mathrm{Sl}(2;\mathbb{R})$ rotated solution will depend on the
original physical parameters
$\hat{q}_{0},\hat{q}_{1},\Im\mathfrak{m}\, \hat{\lambda}_{\infty}$
plus the parameters of the $\mathrm{Sl}(2;\mathbb{R})$ transformation
$a,b,c,d$ (only 3 of which are independent). We have to determine
$\hat{q}_{0},\hat{q}_{1},\Im\mathfrak{m}\,
\hat{\lambda}_{\infty},a,b,c,d$ in terms of the final physical
parameters to write the rotated solution in terms of its own physical
parameters only.

$\mathrm{Sl}(2;\mathbb{R})$ acts on the charge vector through the matrix
Eq.~(\ref{eq:sl2embeddedinsp4}) so

\begin{equation}
\label{eq:rotatingthecharges}
\left(
  \begin{array}{c}
p^{0} \\ p^{1} \\ q_{0} \\ q_{1} \\    
  \end{array}
\right)
=
\left(
  \begin{array}{cccc}
   d &   & -c &   \\
     & a &    & b \\
  -b &   & a  &   \\
     & c &    & d \\ 
  \end{array}
\right)  
\left(
  \begin{array}{c}
0 \\ 0 \\ \hat{q}_{0} \\ \hat{q}_{1} \\    
  \end{array}
\right)
=
\left(
  \begin{array}{r}
-c \hat{q}_{0} \\ b\hat{q}_{1} \\ a\hat{q}_{0} \\ d\hat{q}_{1} \\    
  \end{array}
\right)\, .
\end{equation}

From these relations we determine $a,b,c,d$ in terms of the final and
original charges:

\begin{equation}
\label{eq:abcd}
a= q_{0}/\hat{q}_{0}\, ,
\hspace{.5cm} 
b= p^{1}/\hat{q}_{1}\, ,
\hspace{.5cm}
c= -p^{0}/\hat{q}_{0}\, ,
\hspace{.5cm}  
d= q_{1}/\hat{q}_{1}\, .
\end{equation}

On the other hand, from the transformation rule
Eq.~(\ref{eq:finitesl2rtransformations}) we get

\begin{equation}
\label{eq:esasecuaciones}
\Re\mathfrak{e}\, \lambda_{\infty} 
= 
\frac{ bd +ac (\Im\mathfrak{m}\, \hat{\lambda}_{\infty})^{2}}{d^{2} 
+c^{2} (\Im\mathfrak{m}\, \hat{\lambda}_{\infty})^{2}}\, ,  
\hspace{1cm}
\Im\mathfrak{m}\, \lambda_{\infty} 
= 
\frac{\Im\mathfrak{m}\,  \hat{\lambda}_{\infty}}{d^{2} 
+c^{2} (\Im\mathfrak{m}\, \hat{\lambda}_{\infty})^{2}}\, ,  
\end{equation}

\noindent
and replacing in these relations the transformation parameters
$a,b,c,d$ by the values in Eq.~(\ref{eq:abcd}), we get 2 equations
that relate the 3 original to the 6 final physical parameters:

\begin{align}
\label{eq:axidilatonGen} p^{0}q_{0} (\hat{q}_{1})^{2}  (\Im\mathfrak{m}\,
\hat{\lambda}_{\infty})^{2}
+
\frac{\Re\mathfrak{e}\, \lambda_{\infty}}{\Im\mathfrak{m}\, \lambda_{\infty}}
(\hat{q}_{0}\hat{q}_{1})^{2} \Im\mathfrak{m}\, \hat{\lambda}_{\infty}
-
p^{1}q_{1} (\hat{q}_{0})^{2} 
& = 
0\, ,  
\\
& \nonumber \\
\label{eq:axidilatonGen2}\Im\mathfrak{m}\, \lambda_{\infty}
(p^{0})^{2} (\hat{q}_{1})^{2}  (\Im\mathfrak{m}\, \hat{\lambda}_{\infty})^{2}
-
(\hat{q}_{0}\hat{q}_{1})^{2}\Im\mathfrak{m}\, \hat{\lambda}_{\infty} 
+\Im\mathfrak{m}\, \lambda_{\infty}
(q_{1})^{2} (\hat{q}_{0})^{2} 
& = 
0\, .  
\end{align}

\noindent
The invariance of $\mathsf{W}$ implies that

\begin{equation}
\label{eq:hatqversusq}
\hat{q}_{0}\hat{q}_{1}
= 
p^{0}p^{1}+q_{0}q_{1} 
\, ,
\end{equation}

\noindent
and allows us to eliminate $\hat{q}_{1}$ from the above two
equations. We can solve \eqref{eq:axidilatonGen} and \eqref{eq:axidilatonGen2} for $\Im\mathfrak{m}\,
\hat{\lambda}_{\infty}$ as a function of the 6 final physical
parameters and $\hat{q}_{0}$ and, for both equations, we find
$\Im\mathfrak{m}\, \hat{\lambda}_{\infty} \hat{q}_{0}^{-2}$ as a
function of those 6 parameters:

\begin{equation}
  \Im\mathfrak{m}\, \hat{\lambda}_{\infty} \hat{q}_{0}^{-2} = 
f_{1}(\mathcal{Q}, \lambda_{\infty})\, ,
\hspace{1cm}  
  \Im\mathfrak{m}\, \hat{\lambda}_{\infty} \hat{q}_{0}^{-2} = 
f_{2}(\mathcal{Q}, \lambda_{\infty})\, .
\end{equation}

\noindent
The consistency condition $f_{1}(\mathcal{Q}, \lambda_{\infty}) =
f_{2}(\mathcal{Q}, \lambda_{\infty})$ determines one of the two final
real moduli as a complicated function of the final charges. In other
words: the final solution cannot have 6 independent physical
parameters, which implies that the original solution is not a
generating solution.

On top of this, there seems to be another problem: we cannot solve
separately the 3 original physical parameters in terms of the 6 final
ones. ``Fortunately'' only the combination $\Im\mathfrak{m}\,
\hat{\lambda}_{\infty} \hat{q}_{0}^{-2}$ appears in the rotated
solution or, equivalently, in the $H^{M}$ variables. Using
Eqs.~(\ref{eq:rotatingthecharges},\ref{eq:abcd}) and
(\ref{eq:hatqversusq}) we find the these are given by

\begin{equation}
\label{eq:rotatingtheHs}
H^{M} = A^{M} -\tfrac{1}{\sqrt{2}}\mathcal{Q}^{M} \tau\, ,
\hspace{1cm}
\left(
  \begin{array}{c}
A^{0} \\ A^{1} \\ A_{0} \\ A_{1} \\    
  \end{array}
\right)
=
\left(
  \begin{array}{c}
\frac{s}{\sqrt{2}} p^{0} 
(\Im\mathfrak{m}\, \hat{\lambda}_{\infty} \hat{q}_{0}^{-2})^{1/2}  
\\ 
\frac{s}{\sqrt{2}} p^{1}(p^{0}p^{1}+q_{0}q_{1})^{-1} 
(\Im\mathfrak{m}\, \hat{\lambda}_{\infty} \hat{q}_{0}^{-2})^{-1/2}  
\\ 
\frac{s}{\sqrt{2}} q_{0} 
(\Im\mathfrak{m}\, \hat{\lambda}_{\infty} \hat{q}_{0}^{-2})^{1/2}  
\\ 
\frac{s}{\sqrt{2}} q_{1}(p^{0}p^{1}+q_{0}q_{1})^{-1} 
(\Im\mathfrak{m}\, \hat{\lambda}_{\infty} \hat{q}_{0}^{-2})^{-1/2}  
\\    
  \end{array}
\right)\, ,
\end{equation}

In the supersymmetric case we know that we can construct a new
solution which has, on top of the two non-trivial harmonic functions,
two constant ones. If we write all of them in the form

\begin{equation}
\hat{H}^{M}
=
\hat{A}^{M} -\tfrac{1}{\sqrt{2}}\hat{\mathcal{Q}}^{M} \tau\, ,
\end{equation}

\noindent
then $(\hat{\mathcal{Q}}^{M})^{T} =(0, 0, \hat{q}_{0}, \hat{q}_{1})^{T}$ and,
according to the general results of Ref.~\cite{Galli:2012pt},

\begin{equation}
  (\hat{A}^{M})
  =
  \frac{1}{\sqrt{2\Im\mathfrak{m}\, \hat{\lambda}_{\infty}}}
  \Im\mathfrak{m}
  \left\{
    \frac{\hat{q}_{1} \hat{\lambda}^{*}_{\infty} -i \hat{q}_{0}}{|\hat{q}_{1} \hat{\lambda}^{*}_{\infty} -i \hat{q}_{0}|}
    \left(
  \begin{array}{c}
i \\ \hat{\lambda}_{\infty} \\ -i\hat{\lambda}_{\infty} \\ 1 \\    
  \end{array}
\right)
\right\}\, .
\end{equation}

This solution has two independent charges at any generic point in
moduli space and should be a generating solution. The difference with
the previous case is that, instead of the
Eqs.~(\ref{eq:esasecuaciones}), we can invert
(\ref{eq:finitesl2rtransformations}) and use Eqs.~(\ref{eq:abcd}) and
(\ref{eq:hatqversusq}) to get two independent real equations that do
not lead to constraints in the final physical parameters:

\begin{equation}
\hat{\lambda}_{\infty}\hat{q}_{0}^{-2} 
= 
\frac{1}{(p^{0}p^{1}+q_{0}q_{1})}
\frac{q_{1} \lambda_{\infty} -p^{1}}{p^{0}\lambda_{\infty} +q_{0}}\, .
\end{equation}

The only combinations of the 4 original physical parameters that
appear in the rotated solution are precisely the real and imaginary
parts of $\hat{\lambda}_{\infty}\hat{q}_{0}^{-2}$ and we obtain a
solution with 6 independent physical parameters.


\subsection{2-charge solutions of the $t^{3}$ model}
\label{sec-2charget3}


Again, the minimal number of non-vanishing charges that a regular,
extremal, black hole of this model can have is two. A choice of charge
vector that leads to regular supersymmetric and non-supersymmetric
black holes is $(0,p^{1},q_{0},0)^{T}$. In the supersymmetric case,
the coefficient of $-\tfrac{1}{\sqrt{2}}\tau$ in $H^{M}$ (that we call
attractor in the context of this formalism) is given by

\begin{equation}
(B^{M})=(\mathcal{Q}^{M}) 
=
\left(
  \begin{array}{c}
  0 \\ p^{1} \\ q_{0} \\ 0 \\  
  \end{array}
\right)\, ,
\end{equation}

\noindent
and in the non-supersymmetric one, by

\begin{equation}
(B^{M})
=
\left(
  \begin{array}{c}
  0 \\ p^{1} \\ -q_{0} \\ 0 \\  
  \end{array}
\right)\, .
\end{equation}

In order to see if these charge configurations lead to generating
solutions, we study the values of the invariants.  For cubic models
with prepotential of the form

\begin{equation}
\mathcal{F}
=
\tfrac{1}{3!}d_{ijk}
\frac{\mathcal{X}^{i}\mathcal{X}^{j}\mathcal{X}^{k}}{\mathcal{X}^{0}}\, ,
\end{equation}

\noindent
one has $\mathcal{C}_{ijk} = e^{\mathcal{K}}d_{ijk}$. The prepotential of the
$t^{3}$ model is given in Eq.~(\ref{eq:t3prepotential}) and has $d_{111}=-5$
so $\mathcal{C}_{ttt} = \tfrac{3}{4}(\Im\mathfrak{m}t)^{-3}$. For this model
it can be proven that only three invariants are independent and that the other
two can be written as a their combination. Specifically, one finds that
\cite{Ceresole:2009iy}

\begin{eqnarray}
i_{4} 
& = &
-\sqrt{\tfrac{4}{27}i_{2}^{3}i_{1}-i_{3}^{2}}\, ,
\\
i_{5}
& = & 
\tfrac{3}{4}i_{2}^{2}\, ,
\end{eqnarray}

\noindent
and we can take, as independent basis of invariants $i_{1},i_{2}$ and $i_{3}$
(which we can replace by $J_{4}$). 

Now let us evaluate these invariants for the solutions with charge vector
$(0,p^{1},q_{0},0)^{T}$. The result is

\begin{eqnarray}
i_{1}
& = & 
\frac{3}{20 (\Im\mathfrak{m}\, t_{\infty})^{3}}
\left|-\tfrac{5}{2}p^{1}t_{\infty}^{2}-q_{0}\right|^{2}\, ,
\\
& & \nonumber \\
i_{2}
& = & 
\frac{1}{20(\Im\mathfrak{m}\, t_{\infty})^{3}}
\left|-\tfrac{5}{2}p^{1}t_{\infty}(t_{\infty}+2t_{\infty}^{*})-3q_{0}\right|^{2}\, ,
\\
& & \nonumber \\
i_{3}
& = & 
-\frac{1}{75(\Im\mathfrak{m}\, t_{\infty})^{6}}
\Re\mathfrak{e}\, 
\left\{
-\tfrac{i}{8}\left(-\tfrac{5}{2}p^{1}t_{\infty}^{2}-q_{0}\right) 
\left[-\tfrac{5}{2}p^{1}t_{\infty}(t_{\infty}+2t_{\infty}^{*})-3q_{0} \right]^{3}
\right\}\, ,
\end{eqnarray}

\noindent
and it is easy to see that if $\Re\mathfrak{e}\, t_{\infty}=0$ (the
\textit{axion-free} case) they simplify to

\begin{eqnarray}
i_{1}
& = & 
\frac{3}{20 (\Im\mathfrak{m}\, t_{\infty})^{3}}
\left[\tfrac{5}{2}p^{1}(\Im\mathfrak{m}\, t_{\infty})^{2}-q_{0}\right]^{2}\, ,
\\
& & \nonumber \\
i_{2}
& = & 
\frac{1}{20(\Im\mathfrak{m}\, t_{\infty})^{3}}
\left[\tfrac{5}{2}p^{1}(\Im\mathfrak{m}\, t_{\infty})^{2}+3q_{0}\right]^{2}\, ,
\\
& & \nonumber \\
i_{3}
& = & 0\, ,
\end{eqnarray}

We see then that in the axion-free case only two invariant are
independent and according to the argument in \cite{Bertolini:1999je}
the solutions cannot be seed (generating) solutions.

It is necessary to have $\Re\mathfrak{e}\ t\neq 0$ for the the three
invariants $i_{1},i_{2},i_{3}\neq0$ to be independent from each other
and the two-charge solution to be a generating solution.


\end{document}